\documentclass[twocolumn, superscriptaddress, amsmath,amssymb, prc aps]{revtex4-1}

\usepackage{romanbar}
\usepackage{graphicx}
\usepackage{dcolumn}
\usepackage{bm}
\usepackage{array}
\usepackage{graphicx}  
\usepackage{siunitx}  
\usepackage{epsfig} 
\usepackage{subfigure}
\usepackage{lineno}  
\usepackage{amsmath,amssymb}
\usepackage{comment}
\usepackage{multirow}
\usepackage{dcolumn}
\usepackage{bm}
\usepackage{setspace}
\includecomment{printrobustnotes}
\includecomment{printallnotes}
\usepackage{xspace}
\usepackage{color}
\usepackage{hyperref} 
\begin{document}

\title{Search for solar bosonic dark matter annual modulation with COSINE-100}
\author{G.~Adhikari}
\affiliation{Department of Physics and Wright Laboratory, Yale University, New Haven, CT 06520, USA}
\author{N.~Carlin}
\affiliation{Physics Institute, University of S\~{a}o Paulo, 05508-090, S\~{a}o Paulo, Brazil}
\author{J.~J.~Choi}
\affiliation{Department of Physics and Astronomy, Seoul National University, Seoul 08826, Republic of Korea} 
\affiliation{Center for Underground Physics, Institute for Basic Science (IBS), Daejeon 34126, Republic of Korea}
\author{S.~Choi}
\affiliation{Department of Physics and Astronomy, Seoul National University, Seoul 08826, Republic of Korea} 
\author{A.~C.~Ezeribe}
\affiliation{Department of Physics and Astronomy, University of Sheffield, Sheffield S3 7RH, United Kingdom}
\author{L.~E.~Fran{\c c}a}
\email{luis.eduardo.franca@usp.br}
\affiliation{Physics Institute, University of S\~{a}o Paulo, 05508-090, S\~{a}o Paulo, Brazil}
\author{C.~Ha}
\affiliation{Department of Physics, Chung-Ang University, Seoul 06973, Republic of Korea}
\author{I.~S.~Hahn}
\affiliation{Department of Science Education, Ewha Womans University, Seoul 03760, Republic of Korea} 
\affiliation{Center for Exotic Nuclear Studies, Institute for Basic Science (IBS), Daejeon 34126, Republic of Korea}
\affiliation{IBS School, University of Science and Technology (UST), Daejeon 34113, Republic of Korea}
\author{S.~J.~Hollick}
\affiliation{Department of Physics and Wright Laboratory, Yale University, New Haven, CT 06520, USA}
\author{E.~J.~Jeon}
\affiliation{Center for Underground Physics, Institute for Basic Science (IBS), Daejeon 34126, Republic of Korea}
\author{J.~H.~Jo}
\affiliation{Department of Physics and Wright Laboratory, Yale University, New Haven, CT 06520, USA}
\author{H.~W.~Joo}
\affiliation{Department of Physics and Astronomy, Seoul National University, Seoul 08826, Republic of Korea} 
\author{W.~G.~Kang}
\affiliation{Center for Underground Physics, Institute for Basic Science (IBS), Daejeon 34126, Republic of Korea}
\author{M.~Kauer}
\affiliation{Department of Physics and Wisconsin IceCube Particle Astrophysics Center, University of Wisconsin-Madison, Madison, WI 53706, USA}
\author{B.~H.~Kim}
\affiliation{Center for Underground Physics, Institute for Basic Science (IBS), Daejeon 34126, Republic of Korea}
\author{H.~J.~Kim}
\affiliation{Department of Physics, Kyungpook National University, Daegu 41566, Republic of Korea}
\author{J.~Kim}
\affiliation{Department of Physics, Chung-Ang University, Seoul 06973, Republic of Korea}
\author{K.~W.~Kim}
\affiliation{Center for Underground Physics, Institute for Basic Science (IBS), Daejeon 34126, Republic of Korea}
\author{S.~H.~Kim}
\affiliation{Center for Underground Physics, Institute for Basic Science (IBS), Daejeon 34126, Republic of Korea}
\author{S.~K.~Kim}
\affiliation{Department of Physics and Astronomy, Seoul National University, Seoul 08826, Republic of Korea}
\author{W.~K.~Kim}
\affiliation{IBS School, University of Science and Technology (UST), Daejeon 34113, Republic of Korea}
\affiliation{Center for Underground Physics, Institute for Basic Science (IBS), Daejeon 34126, Republic of Korea}
\author{Y.~D.~Kim}
\affiliation{Center for Underground Physics, Institute for Basic Science (IBS), Daejeon 34126, Republic of Korea}
\affiliation{Department of Physics, Sejong University, Seoul 05006, Republic of Korea}
\affiliation{IBS School, University of Science and Technology (UST), Daejeon 34113, Republic of Korea}
\author{Y.~H.~Kim}
\affiliation{Center for Underground Physics, Institute for Basic Science (IBS), Daejeon 34126, Republic of Korea}
\affiliation{Korea Research Institute of Standards and Science, Daejeon 34113, Republic of Korea}
\affiliation{IBS School, University of Science and Technology (UST), Daejeon 34113, Republic of Korea}
\author{Y.~J.~Ko}
\affiliation{Center for Underground Physics, Institute for Basic Science (IBS), Daejeon 34126, Republic of Korea}
\author{D.~H.~Lee}
\affiliation{Department of Physics, Kyungpook National University, Daegu 41566, Republic of Korea}
\author{E.~K.~Lee}
\affiliation{Center for Underground Physics, Institute for Basic Science (IBS), Daejeon 34126, Republic of Korea}
\author{H.~Lee}
\affiliation{IBS School, University of Science and Technology (UST), Daejeon 34113, Republic of Korea}
\affiliation{Center for Underground Physics, Institute for Basic Science (IBS), Daejeon 34126, Republic of Korea}
\author{H.~S.~Lee}
\affiliation{Center for Underground Physics, Institute for Basic Science (IBS), Daejeon 34126, Republic of Korea}
\affiliation{IBS School, University of Science and Technology (UST), Daejeon 34113, Republic of Korea}
\author{H.~Y.~Lee}
\affiliation{Center for Underground Physics, Institute for Basic Science (IBS), Daejeon 34126, Republic of Korea}
\author{I.~S.~Lee}
\affiliation{Center for Underground Physics, Institute for Basic Science (IBS), Daejeon 34126, Republic of Korea}
\author{J.~Lee}
\affiliation{Center for Underground Physics, Institute for Basic Science (IBS), Daejeon 34126, Republic of Korea}
\author{J.~Y.~Lee}
\affiliation{Department of Physics, Kyungpook National University, Daegu 41566, Republic of Korea}
\author{M.~H.~Lee}
\affiliation{Center for Underground Physics, Institute for Basic Science (IBS), Daejeon 34126, Republic of Korea}
\affiliation{IBS School, University of Science and Technology (UST), Daejeon 34113, Republic of Korea}
\author{S.~H.~Lee}
\affiliation{IBS School, University of Science and Technology (UST), Daejeon 34113, Republic of Korea}
\affiliation{Center for Underground Physics, Institute for Basic Science (IBS), Daejeon 34126, Republic of Korea}
\author{S.~M.~Lee}
\affiliation{Department of Physics and Astronomy, Seoul National University, Seoul 08826, Republic of Korea} 
\author{Y.~J.~Lee}
\affiliation{Department of Physics, Chung-Ang University, Seoul 06973, Republic of Korea}
\author{D.~S.~Leonard}
\affiliation{Center for Underground Physics, Institute for Basic Science (IBS), Daejeon 34126, Republic of Korea}
\author{N.~T.~Luan}
\affiliation{Department of Physics, Kyungpook National University, Daegu 41566, Republic of Korea}
\author{B.~B.~Manzato}
\affiliation{Physics Institute, University of S\~{a}o Paulo, 05508-090, S\~{a}o Paulo, Brazil}
\author{R.~H.~Maruyama}
\affiliation{Department of Physics and Wright Laboratory, Yale University, New Haven, CT 06520, USA}
\author{R.~J.~Neal}
\affiliation{Department of Physics and Astronomy, University of Sheffield, Sheffield S3 7RH, United Kingdom}
\author{J.~A.~Nikkel}
\affiliation{Department of Physics and Wright Laboratory, Yale University, New Haven, CT 06520, USA}
\author{S.~L.~Olsen}
\affiliation{Center for Underground Physics, Institute for Basic Science (IBS), Daejeon 34126, Republic of Korea}
\author{B.~J.~Park}
\affiliation{IBS School, University of Science and Technology (UST), Daejeon 34113, Republic of Korea}
\affiliation{Center for Underground Physics, Institute for Basic Science (IBS), Daejeon 34126, Republic of Korea}
\author{H.~K.~Park}
\affiliation{Department of Accelerator Science, Korea University, Sejong 30019, Republic of Korea}
\author{H.~S.~Park}
\affiliation{Korea Research Institute of Standards and Science, Daejeon 34113, Republic of Korea}
\author{K.~S.~Park}
\affiliation{Center for Underground Physics, Institute for Basic Science (IBS), Daejeon 34126, Republic of Korea}
\author{S.~D.~Park}
\affiliation{Department of Physics, Kyungpook National University, Daegu 41566, Republic of Korea}
\author{R.~L.~C.~Pitta}
\affiliation{Physics Institute, University of S\~{a}o Paulo, 05508-090, S\~{a}o Paulo, Brazil}
\author{H.~Prihtiadi}
\affiliation{Center for Underground Physics, Institute for Basic Science (IBS), Daejeon 34126, Republic of Korea}
\author{S.~J.~Ra}
\affiliation{Center for Underground Physics, Institute for Basic Science (IBS), Daejeon 34126, Republic of Korea}
\author{C.~Rott}
\affiliation{Department of Physics, Sungkyunkwan University, Suwon 16419, Republic of Korea}
\affiliation{Department of Physics and Astronomy, University of Utah, Salt Lake City, UT 84112, USA}
\author{K.~A.~Shin}
\affiliation{Center for Underground Physics, Institute for Basic Science (IBS), Daejeon 34126, Republic of Korea}
\author{D.~F.~F.~S. Cavalcante}
\affiliation{Physics Institute, University of S\~{a}o Paulo, 05508-090, S\~{a}o Paulo, Brazil}
\author{A.~Scarff}
\affiliation{Department of Physics and Astronomy, University of Sheffield, Sheffield S3 7RH, United Kingdom}
\author{N.~J.~C.~Spooner}
\affiliation{Department of Physics and Astronomy, University of Sheffield, Sheffield S3 7RH, United Kingdom}
\author{W.~G.~Thompson}
\affiliation{Department of Physics and Wright Laboratory, Yale University, New Haven, CT 06520, USA}
\author{L.~Yang}
\affiliation{Department of Physics, University of California San Diego, La Jolla, CA 92093, USA}
\author{G.~H.~Yu}
\affiliation{Department of Physics, Sungkyunkwan University, Suwon 16419, Republic of Korea}
\affiliation{Center for Underground Physics, Institute for Basic Science (IBS), Daejeon 34126, Republic of Korea}
\collaboration{COSINE-100 Collaboration}
\date{\today}

\begin{abstract}
    We present results from a search for solar bosonic dark matter using the annual modulation method with the COSINE-100 experiment. The results were interpreted considering three dark sector bosons models: solar dark photon; DFSZ and KSVZ solar axion; and Kaluza-Klein solar axion. No modulation signal that is compatible with the expected from the models was found from a data-set of 2.82 yr, using 61.3 kg of NaI(Tl) crystals. Therefore, we set a 90\% confidence level upper limits for each of the three models studied. For the solar dark photon model, the most stringent mixing parameter upper limit is $1.61 \times 10^{-14}$ for dark photons with a mass of 215 eV. For the DFSZ and KSVZ solar axion, and the Kaluza-Klein axion models, the upper limits exclude axion-electron couplings, $g_{ae}$, above $1.61 \times 10^{-11}$ for axion mass below 0.2 keV; and axion-photon couplings, $g_{a\gamma\gamma}$, above $1.83 \times 10^{-11}$ GeV$^{-1}$ for an axion number density of $4.07 \times 10^{13}$ cm$^{-3}$. This is the first experimental search for solar dark photons and DFSZ and KSVZ solar axions using the annual modulation method. The lower background, higher light yield and reduced threshold of NaI(Tl) crystals of the future COSINE-200 experiment are expected to enhance the sensitivity of the analysis shown in this paper. We show the sensitivities for the three models studied, considering the same search method with COSINE-200.
\end{abstract}

\maketitle

\section{Introduction}\label{sec:introduction}
Although considerable astrophysical and cosmological evidence demonstrates the existence of dark matter \cite{1,2,3,4,5,6,7}, very little is known about its particle properties. For already some time, the main dark matter candidate that has been searched for by direct detection experiments is the Weakly Interacting Massive Particle (WIMP) \cite{8}, and, to date, no convincing evidence for dark matter has been reported \cite{9}. DAMA/LIBRA is the only experiment to claim the observation of events caused by dark matter, albeit its results are refuted by several experiments \cite{10,11,12,13,14,15,16,17}. Moreover, with the absense of any signal at colliders, in recent years the dark matter searches have considered new scenarios that have non-WIMP-like dark matter particle candidates. Bosonic dark matter is considered in different scenarios, including dark photons, which could mediate interactions between Standard Model particles and the dark sector \cite{18}, and axions, which could provide a solution to the CP issue in the QCD \cite{19}. Even though cosmological and astrophysical observations determine stringent upper limits for bosonic dark matter \cite{20,21,22,23,24}, with the sensitivity enhancement of modern dark matter detectors, some experiments are already capable of examining unexplored regions in the parameter spaces, attracting the interest to the direct search of these particles. Since they would be created in processes common in stellar interiors, the Sun is a potential source of axions and dark photons on Earth \cite{25}. 

Most experiments study solar bosonic dark matter particles looking for excess events in electronic recoils \cite{26,27,28,29}. However, in experiments with multiple years of data taking, it is also possible to search for an annual modulation in the detectors' event rate that is caused by the variation of the distance between Earth and Sun during the year \cite{30}. Although the search for excess events is a more sensitive procedure, it is often model dependent, while the annual modulation search method is more general, and its results could be interpreted in the context of any solar dark matter model. In particular, we consider three commonly studied solar bosonic dark matter models as guides for this analysis.

In this paper, results for the search of solar bosonic dark matter expected annual modulation in the COSINE-100 experiment are presented. A data-set of 2.82 yr, using 61.3 kg of NaI(Tl) crystals was analyzed. Three different models for solar dark matter were studied: dark photons \cite{25}; Dine-Fischler-Srednicki-Zhitnisky (DFSZ) and Kim-Shifman-Vainshtein-Zakharov (KSVZ) axions \cite{31}; and Kaluza-Klein (KK) axions \cite{32}. This work is the first to present results from the annual modulation search method for solar dark photons and DFSZ and KSVZ solar axions.

The sensitivities for these models that will be provided by the future COSINE-200 experiment are also presented and discussed. The COSINE-200 NaI(Tl) crystals will have approximately 10 times lower background, a lower energy threshold, and higher light yield than the current COSINE-100 crystals, enhancing the detector's sensitivity to the analysis presented in this paper. 

\section{Experiment}
\begin{figure}[!htb]
    \centering
    \includegraphics[scale=0.45, trim = 0mm 0mm 0mm 0mm]{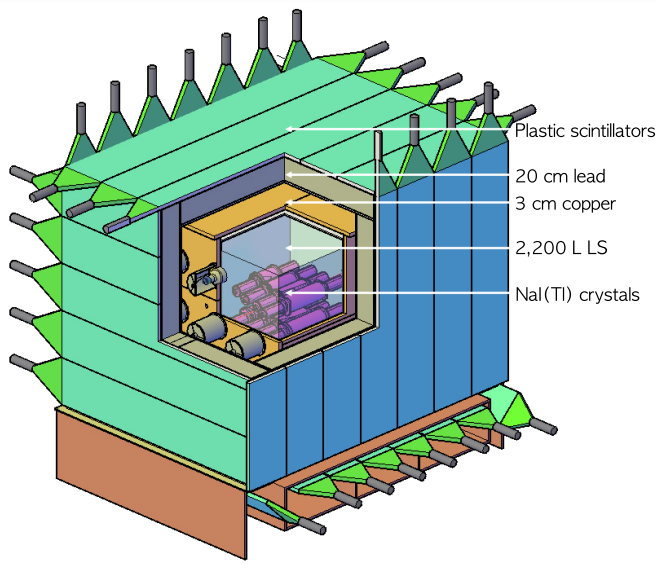}
    \caption{Schematic of COSINE-100 detector, showing the positioning of NaI(Tl) crystals, as well as the passive and active shields \cite{35}.}
    \label{fig:1}
\end{figure}

The COSINE-100 experiment is installed in the A5 tunnel at the Yangyang underground laboratory (Y2L), in South Korea. The main COSINE-100 detectors are 8 ultra-pure NaI(Tl) crystals, with a total of 106 kg, which are immersed in 2200 liters of liquid scintillator (LS) \cite{33}. Surrounding the LS are copper and lead shields that are surrounded by plastic scintillators (PS) that detect cosmic rays that transverse that apparatus \cite{34}. The LS and PS act as active shields, since they are able to detect background radiation present in the detector. All signals from the photomultipliers (PMTs) coupled to these two detectors are processed by 63.5 mega sample per second (MSPS) analog to digital converter (M64ADC) modules. The copper and lead act as passive shields, and reduce external background that could hit the crystals. Figure \ref{fig:1} shows a schematic of the COSINE-100 experiment.

All eight encased crystal detectors are placed on an acrylic table in a 4 $\times$ 2 array. They were developed in cooperation with Alpha Spectra Inc. After their production, the crystals were encapsulated in oxygen free copper tubes, and quartz windows were attached to their end-faces. R12669SEL PMTs from Hamamatsu Photonics are coupled to each end of the crystals. Every PMT has two outputs: the anode, whose signals are used for events with energies below 100 keV, and the dynode, whose signals are used for events with energies up to a few MeVs. Both output signals are sent to pre-amplifiers before being processed by the 500 MSPS fast-ADC (FADC) modules.

If a signal from a certain PMT is above the 6~mV threshold [based on the pulse height], the FADC opens a coincidence window of 200~ns, and waits for a coincident signal from the other PMT of the crystal. If the coincidence happens, the trigger and control board generates a global trigger, and the data from all FADC and M64ADC modules are saved.

Environmental variables in the laboratory room such as temperature, radon and humidity levels are monitored and controlled in real time,
while slow physics parameters related to background rates and detector stabilities, are monitored through the Grafana application \cite{36} by shifters.  

Details of the COSINE-100 components, data acquisition and monitoring systems are described in Refs. \cite{37,38,39}. 

\section{Annual Modulation}
Annual modulation in the event rate measured by the detector is one method of detecting solar dark matter. Just like the expected annual modulation generated by WIMPs from the Galactic halo \cite{40}, this search method has a benefit of being independent of the nature of the interaction between dark matter and the detector. Also, the background event rate is not expected to show a modulation behavior, creating a reliable way to distinguish between dark matter and background events. Even though the background does exhibit a time varying behaviour, as detailed in Sec. \ref{sec:analB}, a correct treatment of each background component should not create a bias in the modulation results of the analysis. 

The expected annual modulation has its origin in the elliptical orbit of the Earth. If the Sun emits dark matter particles, the expected flux on Earth should be higher when the Earth is at the perihelion, and lower when the Earth is at the aphelion. During the year, since there is a variation in the distance between Earth and Sun, the flux should also vary in the same way, leading to an annual modulation in the measured event rate.

The solar dark photons and solar DFSZ and KSVZ axions models suggest the measured event rate is proportional to $d^{-2}$, while the solar KK axions model suggest it is proportional to $d^{-4}$, where $d$ is the distance between the Earth and the Sun, which can be written as:

\begin{equation}\label{eq:1}
    d(t) = a \Big(1 + e\,\rm{cos}\,\dfrac{2\,\pi\,(t-t_0)}{T}\Big)   
\end{equation}

\noindent where $a = 1.496 \times 10^{11}$ m is the semi-major axis of Earth's orbit, $e = 0.0167$ is Earth's orbit eccentricity, $t_0 = 3$ days is the phase related to the date of Earth's perihelion, and $T = 1$ sidereal year is the period. 

Assuming $R \propto d^{-2}$, where $R$ is the event rate, its expression can be written as:
\begin{equation}\label{eq:2}
    R = C \times \Big(1 + e\,\rm{cos}\,\dfrac{2\,\pi\,(t-t_0)}{T}\Big)^{-2}
\end{equation}

Expanding this expression up to second order since $e\,\rm{cos}\,\dfrac{2\,\pi\,(t-t_0)}{T} << 1$, the event rate is expected to be:
\begin{equation}\label{eq:3}
    \begin{split}
    R \approx R_{avg} + 2\,e\,R_{avg}\;& \Big(\rm{cos}\,\frac{2\pi(t-t_{0})}{T} + \\*
    & \dfrac{3}{2}\,e\,\rm{cos}^2\,\frac{2\pi(t-t_{0})}{T}\,\Big)
    \end{split}
\end{equation}
    
\noindent where $R_{avg}$ is the event rate when the distance between Earth and the Sun is $d=a=1.496 \times 10^{11}$ m.

In this case, the annual modulation in the event rate will follow:
\begin{equation}\label{eq:4}
    A_{-2}=A\;\Big(\rm{cos}\,\frac{2\pi(t-t_{0})}{T}+\frac{3}{2}\,e\,\rm{cos}^2\,\frac{2\pi(t-t_{0})}{T}\Big) 
\end{equation}

\noindent where $A$ is the amplitude.

Assuming $R \propto d^{-4}$, and also expanding up to second order, the event rate is expected to be: 
\begin{equation}\label{eq:5}
    \begin{split}
    R \approx R_{avg}+4\,e\,R_{avg}\;& \Big(\rm{cos}\,\frac{2\pi(t-t_{0})}{T}+ \\*
    & \dfrac{5}{2}\,e\,\rm{cos}^2\,\frac{2\pi(t-t_{0})}{T}\,\Big)
    \end{split}
\end{equation}
    
In this case, the annual modulation in the event rate should be:
\begin{equation}\label{eq:6}
    A_{-4}=A\;\Big(\rm{cos}\,\frac{2\pi(t-t_{0})}{T}+\frac{5}{2}\,e\,\rm{cos}^2\,\frac{2\pi(t-t_{0})}{T}\Big) 
\end{equation}

\noindent where $A$ is the amplitude.

The $R_{avg}$ value depends on the parameters of each studied model, e.g., on the dark photon or axion mass, as well as the mixing parameter ($\epsilon$), the axion-electron coupling constant, or the axion-photon coupling constant. From the observed amplitude of the expected modulation, it would be possible to determine the parameters of the model considered. 

\section{Dark Photons}
Dark photons have been the subject of many recent intense theoretical studies and experimental searches. This is because they naturally occur in some simple extensions of the Standard Model, and could help solve the hierarchy problem or explain the anomalous muon magnetic moment \cite{41,42}. In string theory, the prediction of dark matter candidates often conjectures the existence of dark photons\cite{43}.

The dark photon would be a boson that belongs to the dark sector, and could be the mediator of an interaction between dark matter particles, or between the standard model and the dark sector \cite{44}. Although it is sterile for standard model interactions, it would have a kinetic mixing term with the ordinary photon. Hence, it would be possible to detect it directly through an effect very similar to the photoelectric effect.

Most of dark photon models consider that its mass originates from the Stueckelberg mechanism \cite{45}. The Higgs \cite{46} and Stueckelberg mechanisms are equivalent in the limit of a dark Higgs mass that is much higher than the dark photon mass. The analysis done in this work is valid if the dark photon mass comes from the Stueckelberg mechanism.

The Lagrangian that describes the system composed of the ordinary photon and the dark photon is \cite{47}:
\begin{equation}\label{eq:7}
    \begin{split}
    \mathcal{L} = - \frac{1}{4}A_{\mu\nu}A^{\mu\nu} - \frac{1}{4}B_{\mu\nu}B^{\mu\nu} & + \frac{m^2_{DP}}{2}B_{\mu}B^{\mu} - \\*
    & \frac{\epsilon}{2}A_{\mu\nu}B^{\mu\nu}
    \end{split}
\end{equation}
\noindent where A is the field associated to the ordinary photon, B is the field associated to the dark photon, $m_{DP}$ is the dark photon mass, and $\epsilon$ is the kinetic coupling parameter.

Considering dark photons with mass below 100 keV, the main source on Earth would be stars, in which a high rate of dark photons is produced by the oscillation of ordinary photons, and should depend on the dark photon mass, as well as the stars' composition and temperature \cite{48}. Specifically on Earth, the main source of dark photons would be the Sun.

The production rate of photons inside the Sun can be determined from the solar opacities, which consider processes like the inverse Compton, Bremsstrahlung, and the photoelectric effect, ans depends on the composition and temperature as a function of the radius inside the Sun \cite{25}. The dark photon production rate in the Sun is connected to the oscillation probability of a photon into a dark photon in a homogeneous medium, which depends on the assumed dark photon polarization \cite{47}. Thus, it is necessary to calculate independently the longitudinal and transverse dark photon polarization fluxes on Earth.  

\begin{figure}[!htb]
    \centering
    \includegraphics[scale=0.365, trim = 0mm 10mm 0mm 0mm]{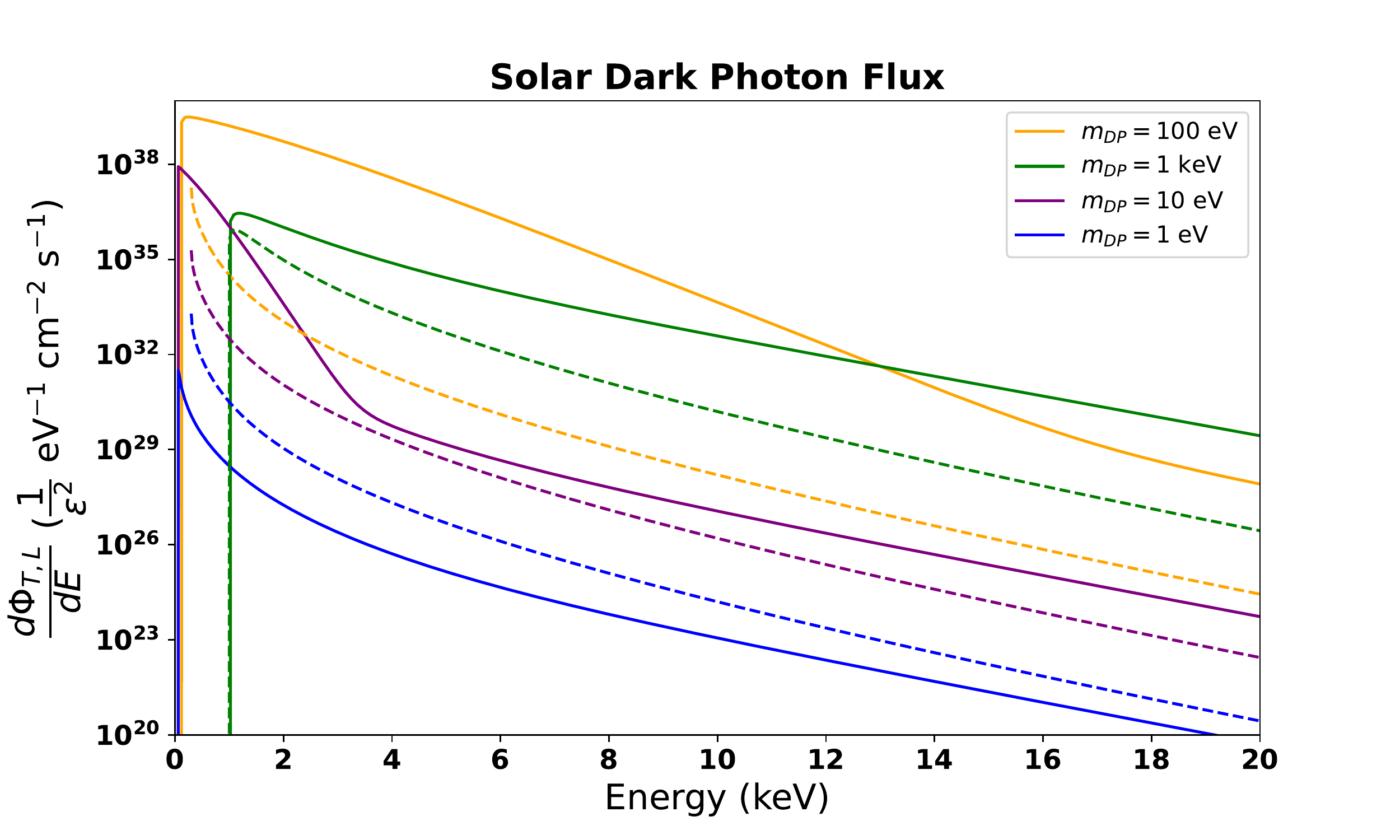}
    \caption{Dark photon fluxes on Earth with longitudinal polarization (dashed lines) and transverse polarization (solid lines), for 100 eV (orange), 1 keV (green), 10 eV (purple), and 1 eV (blue) masses.}
    \label{fig:2}
\end{figure}

According to the procedure described in Refs. \cite{46,47,49}, the dark photon fluxes on Earth with longitudinal and transverse polarizations, and for different masses were calculated, as is shown in Figure \ref{fig:2}. 

For solar dark photons with mass below around 5 eV, the flux due to the longitudinal polarization is dominant, while for higher masses, the flux due to the transverse polarization is dominant. For dark photon masses below around 5 eV and above 316 eV, most dark photons with transverse polarization and energy above 1 keV are generated in the bulk of the Sun \cite{25}. For small masses, the resonant region (when $w_p(r)=m_{DP}$, where $w_p(r)$ is the plasma frequency inside the Sun) is in the outer layers of the Sun, which contributes negligibly to the total flux on Earth. For intermediate masses, the resonant region makes the most important contribution to the flux. Since the highest value of $w_p$ inside the Sun is about 316 eV, for dark photon masses above this value there is no resonant region inside the Sun, and the total flux on Earth decreases.

From the solar dark photons flux on Earth with longitudinal and transverse polarizations, the expected event rate in the COSINE-100 NaI(Tl) crystals can be calculated according to \cite{26}:
\begin{equation}\label{eq:8}
    R_{T,L}=\frac{1}{\rho_{\rm{NaI}}\,v_{DP}}\,\frac{d\Phi_{T,L}}{dE}\,\Gamma_{abs\,T,L}
\end{equation}
    
\noindent where $\rho_{\rm{NaI}}=3.67$ g/cm$^3$ is the NaI density, $v_{DP}$ is the ratio between solar dark photons velocity and the light velocity, $\frac{d\Phi_{T,L}}{dE}$ is the dark photons flux on Earth with longitudinal or transverse polarization, and $\Gamma_{abs\,T,L}$ is the dark photons absorption rate with longitudinal or transverse polarization in NaI(Tl) crystal.

\section{DFSZ and KSVZ Axions}
Another recently studied dark sector boson is the axion, whose existence was not originally proposed as a new dark matter candidate, but, instead, as an explanation for the non-violation of the CP symmetry in the strong interaction. However, its properties make it well suited as a dark matter candidate. 

In hadronic axion models, such as KSVZ, the dominant axion coupling would be to photons, while in non-hadronic models, such as the DFSZ, the dominant coupling would be to electrons (denoted as $g_{ae}$). Considering the axion coupling to electrons, the processes of axion generation are denominated ABC, which include the production by axio-Bremsstrahlung in electron-ion (\ref{eq:9}) and electron-electron collisions (\ref{eq:10}), by Compton (\ref{eq:11}), electronic deexcitation by axions (\ref{eq:12}), and recombination by axions (\ref{eq:13}).

\begin{align}
     e^- + I \longrightarrow e^- + I + a \label{eq:9}\\
     e^- + e^- \longrightarrow e^- + e^- + a \label{eq:10}\\
     \gamma + e^- \longrightarrow e^- + a \label{eq:11}\\
     I^* \longrightarrow I + a \label{eq:12}\\
     e^- + I \longrightarrow I^- + a \label{eq:13}
\end{align}

Considering the axion coupling to photons, axions could be produced by the Primakoff and photocoalescence processes \cite{30}. In hadronic models, axion coupling to photons is dominant, and both of these processes should be considered when studying solar axions. On the other hand, in non-hadronic models, axion coupling to electrons is dominant, and has to be considered in solar axions studies.  

In COSINE-100, the axion detection method that is considered is the axio-electric effect in the NaI(Tl) crystals, which depends only on its coupling to electrons ($g_{ae}$). Therefore, it is possible to consider only the axion electron coupling, with higher sensitivity to non-hadronic models, such as the DFSZ. The $g_{ae}$ coupling depends on the model considered in the analysis. For instance, for DFSZ axions, it is proportional to $\rm{cos}^2 \beta$, where $\rm{cot}\,\beta$ is the ratio between the expected vacuum value of the two Higgs bosons in the model proposed in Ref.\cite{50}. For KSVZ axions, it depends on $E/N$, where $E$ is the coefficient of the electromagnetic anomaly and $N$ is the coefficient of the color anomaly. For the DFSZ model, axions could couple to leptons at tree level, while for the KSVZ model, axions could only couple to leptons at the one-loop level.

The COSINE-100 collaboration has already published the search for solar DFSZ and KSVZ axions with 59.5 days of data, where no excess events were observed \cite{51}. The detection method focused on the axio-electric effect, and the determination of the $g_{ae}$ constant was the objective. Since the interaction between axions and leptons is suppressed in the KSVZ model, the search for DFSZ axions was favored. Low energy events in the crystals were analyzed (from 2 keV to 20 keV), since the solar axion spectrum favors energies from 0.5 keV to 10 keV. The measurements were consistent with the background hypothesis and an upper limit for $g_{ae}$ and the axion mass was determined.

\begin{figure}[!htb]
    \centering
    \includegraphics[scale=0.46, trim = 5mm 10mm 10mm 10mm]{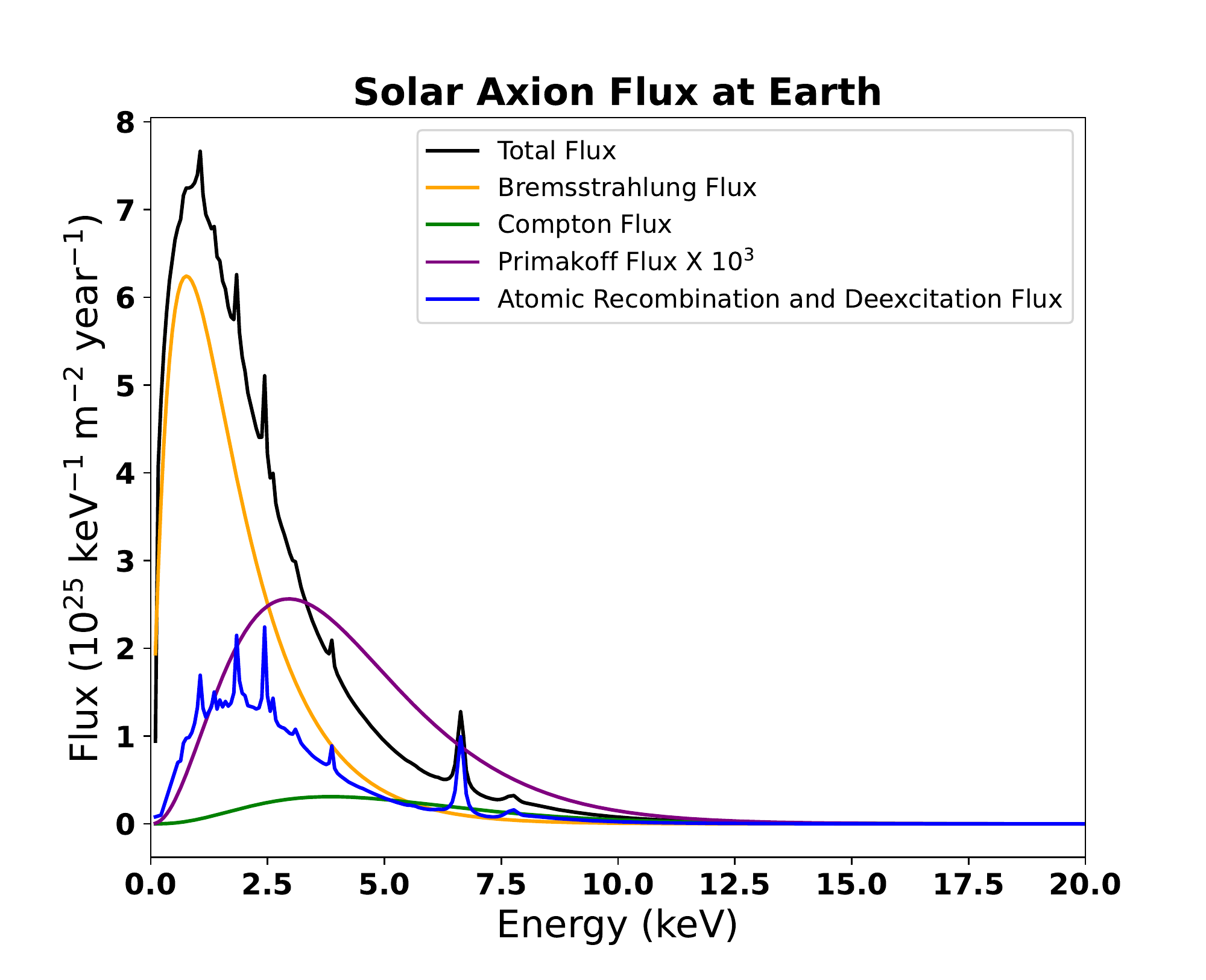}
    \caption{Solar axion flux on Earth considering the axion-Bremsstrahlung process (orange), Compton (green), Primakoff (purple), and electronic recombination and deexcitation by axions (blue). Couplings of $g_{ae}=5.11 \times 10^{-11}$, $m_a=0.01$~eV, and $g_{a\gamma\gamma}=1.02 \times 10^{-10}$~GeV$^{-1}$ were considered, since these are typical values for axions from non-hadronic models. The flux for the Primakoff effect was multiplied by $10^{3}$ for better visualization, and has not been considered in the total flux (black).}
    \label{fig:3}
\end{figure}

Figure \ref{fig:3} shows the solar axion flux on Earth from each of the cited processes in equations \ref{eq:9}-\ref{eq:13}. 

The spikes found in the solar axion flux on Earth are due to the electron binding energies in the atoms present inside the Sun. At the energies of the spikes, there is a resonant axion producion for the electronic deexcitation and recombination (processes \ref{eq:12} and \ref{eq:13}).

The event rate in the NaI(Tl) crystals is given by the following expression \cite{51}:
\begin{equation}\label{eq:14}
    R=\frac{1}{M_{total}}\,\frac{d\Phi_a}{dE_a}\,\Big(\sigma^{\rm{Na}}_{ae}\, N_{\rm{Na}} + \sigma^{\rm{I}}_{ae}\,N_{\rm{I}}\Big)
\end{equation}
    
\noindent where $M_{total}$ is the mass sum of all 5 crystals analyzed, $\frac{d\Phi_a}{dE_a}$ is the solar axion flux on Earth, $\sigma^{\rm{Na}}_{ae}$ and $\sigma^{\rm{I}}_{ae}$ are the axio-electric cross sections for Na and I, respectively, and $N_{\rm{Na}}$ and $N_{\rm{I}}$ are the number of Na and I atoms in the crystals, respectively.

The axio-electric cross sections in the Na and I atoms can be calculated from the photoelectric effect cross sections:
\begin{equation}\label{eq:15}
    \sigma_{ae}=\sigma_{pe}\,\frac{3 E^2_a g^2_{ae}}{16 \pi \alpha m^2_e \beta_a}\,\Big(1-\frac{\beta^{2/3}_a}{3}\Big)
\end{equation}
     
\noindent where $\sigma_{pe}$ is the photoelectric cross section, which can be obtained from Ref. \cite{52}, $E_a$ is the axion energy, $\alpha$ is the fine structure constant, $m_a$ is the axion mass, and $\beta_a$ is the axion velocity, which is defined as $\beta_a=\sqrt{1-\frac{m^2_a}{E^2_a}}$.

\section{Kaluza-Klein Axions}
In the Kaluza Klein (KK) model, axions would propagate in extra dimensions, and could be observed with different mass values. In the KK theory, light particles with masses of O(eV) to O(keV), such as the axions, could propagate in extra dimensions, and would be observed in the conventional 4-dimensions with different mass states. Such mass values would be quantized, and dependent on the number of extra dimensions ``$n$'' \cite{53}. Although the case of $n=1$ is already ruled out since it would cause distortions in the Newtonian gravity, $n=2$ is a possibility, implying a separation of the mass states of 1 eV.  

In hadronic models, in which the coupling between axions and photons $g_{a \gamma \gamma}$ is much higher than $g_{ae}$, and the main processes in solar axion production are the Primakoff effect and photocoalescence, axion propagation in extra dimensions could solve the coronal heating problem in the Sun \cite{54}. Due to the different possible masses in this model, part of solar axions would be produced with velocities smaller than the escape velocity, and would be bound to the solar gravitational well. The number density of KK axions is calculated to be proportional to $r^{-4}$, where $r$ is the distance between the Earth and the Sun. Also, due to the accumulation of KK axions in orbits around the Sun that have existed since the beginning of the solar system, some of these axions would be in orbits that cross the Earth, and could decay into two photons inside detectors. The coupling between axions and photons, which leads to this decay is written as $g_{a \gamma \gamma}$, and the mean decay time is given by:
\begin{equation}\label{eq:16}
    \tau_a = \frac{64\,\pi}{g^2_{a\gamma\gamma}\,m^3_a}
\end{equation}

Hence, another analysis shown in this work is the search for solar KK axions, considering their coupling with photons, with the COSINE-100 experiment. It is expected that the rate of axions decay inside the NaI(Tl) crystals should be proportional to $d^{-4}$, implying in a event rate annual modulation given by expression \ref{eq:5}.   

The expected event rate in the COSINE-100 crystals can be described by:
\begin{equation}\label{eq:17}
    R=\frac{1}{\rho_{\rm{NaI}}}\,\frac{g^2_{a \gamma \gamma}}{64\pi}\,n_a\,m^3_a\,f(m_a)
\end{equation}

\noindent where $\rho_{\rm{NaI}}=3.67$ g/cm$^3$ is the NaI density, $n_a$ is the numerical density of axions at Earth, $m_a$ is the observed axoin mass, and $f(m_a)$ is the axion mass spectrum \cite{55}.

\begin{figure}[!htb]
    \centering
    \includegraphics[scale=0.45, trim = 5mm 10mm 10mm 10mm]{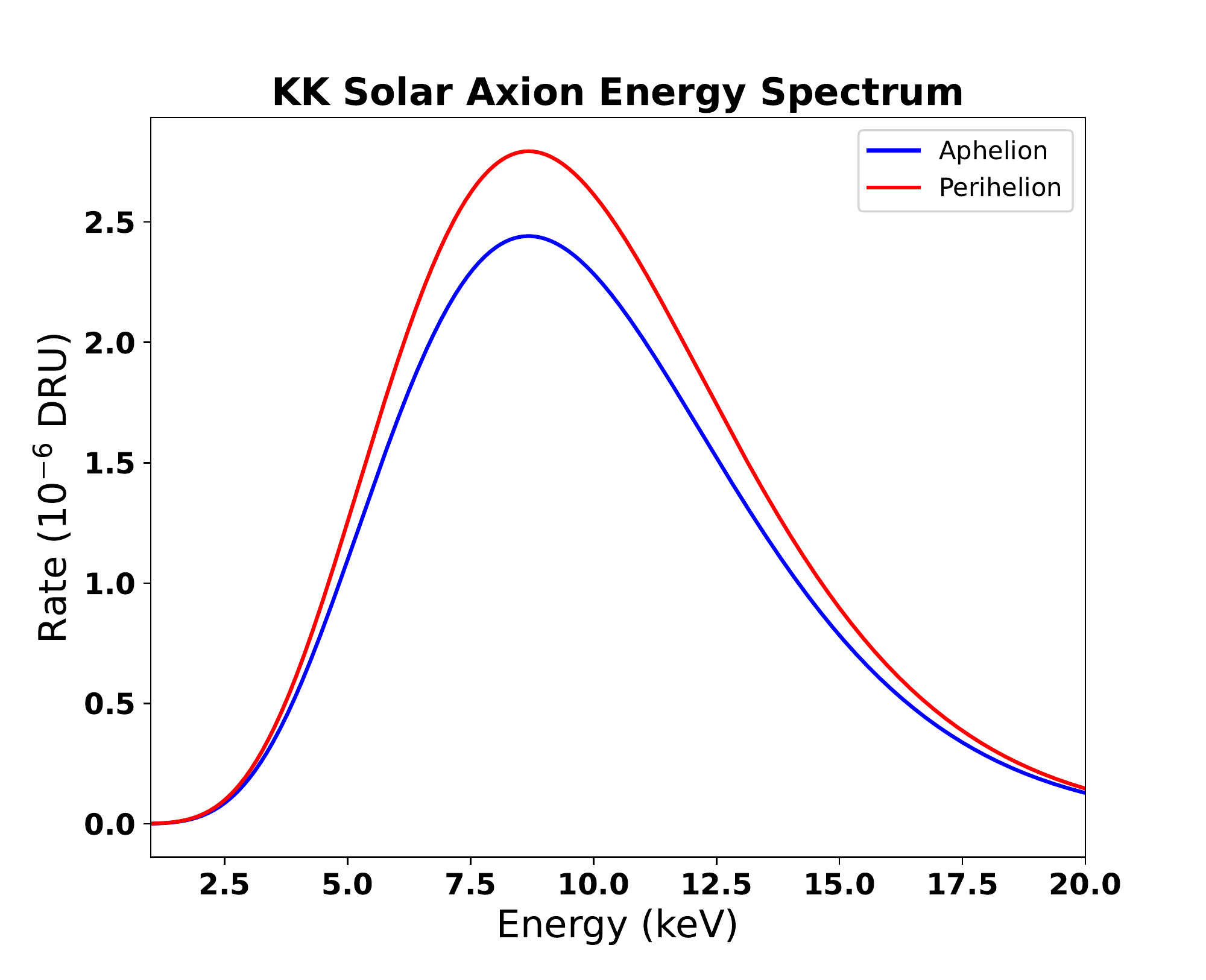}
    \caption{Expected solar KK axions in the COSINE-100 NaI(Tl) crystals when Earth is at aphelion (blue), and perihelion (red), assuming $n_{\rm{aphelion}}=3.81 \times 10^{13}$ m$^{-3}$, $n_{\rm{perihelion}}=4.36 \times 10^{13}$ m$^{-3}$ and $g_{a \gamma \gamma}=9.2 \times 10^{-14}$ GeV$^{-1}$.}
    \label{fig:4}
\end{figure}

Considering the axion numerical density when Earth is at aphelion is $n_{\rm{aphelion}}=3.81 \times 10^{13}$ m$^{-3}$, and when Earth is at perihelion is $n_{\rm{perihelion}}=4.36 \times 10^{13}$ m$^{-3}$, according to Ref. \cite{30}, the expected spectra of the KK axions in the COSINE-100 crystals are as shown in Figure \ref{fig:4}.
    
The event rate has a maximum for energies around 9 keV. Therefore, for this model, the region of interest is between 4 keV and 16 keV, which is advantageous for the COSINE-100 crystals, since their efficiency is near 100\% in this energy region \cite{56}.

\section{Data analysis}
The analysis was performed using data from 5 COSINE-100 crystals, for which the dark photon and axion hypothesis, resulting in an annual modulation, was tested. The mass of the 5 used crystals sum up to 61.3 kg of NaI(Tl). Also, according to the studied bosonic dark matter particles production in the Sun, and to equations \ref{eq:3}-\ref{eq:7}, \ref{eq:8}, \ref{eq:14}, and \ref{eq:17}, the expected event rate and amplitudes in COSINE-100 crystals for 1 keV energy bins, from 1 keV to 20 keV, were calculated. Then, the observed amplitudes from the data were compared with the calculated amplitudes. Based on the background model for each crystal, which is thoroughly studied by the collaboration \cite{57}, the expected modulation fits were performed for each 1 keV energy interval, with the phase fixed in $t_0 = 3$ days and period fixed in $T = 1$ sidereal year. For the dark photons; DFSZ and KSVZ axions; and KK axions, masses from 1 eV to 20 keV; $10^{-5}$ keV to 0.2 keV; and numerical densities from $10^{10}$ m$^{-3}$ to $10^{15}$ m$^{-3}$, respectively, were studied.

\subsection{Event Selection}
Event selections were applied to data in order to reduce or remove background from muons, photons and beta radiation, as well as remove part of the events that originate from noise, which are dominant in energies below 20 keV. 

Only single-hit events were analyzed. Since the interaction probability of dark matter particles with the detector is very small, when an interaction occurs, it should happen in only a single crystal. Also, events generated by muons, or generated in the crystals up to 30 ms after a muon is tagged, were removed, including phosphorescence events activated by muons.

\begin{figure}[!htb]
\begin{center}
\begin{subfigure}(a){}
   \includegraphics[width=0.53\textwidth, trim = 20mm 0mm 20mm 10mm]{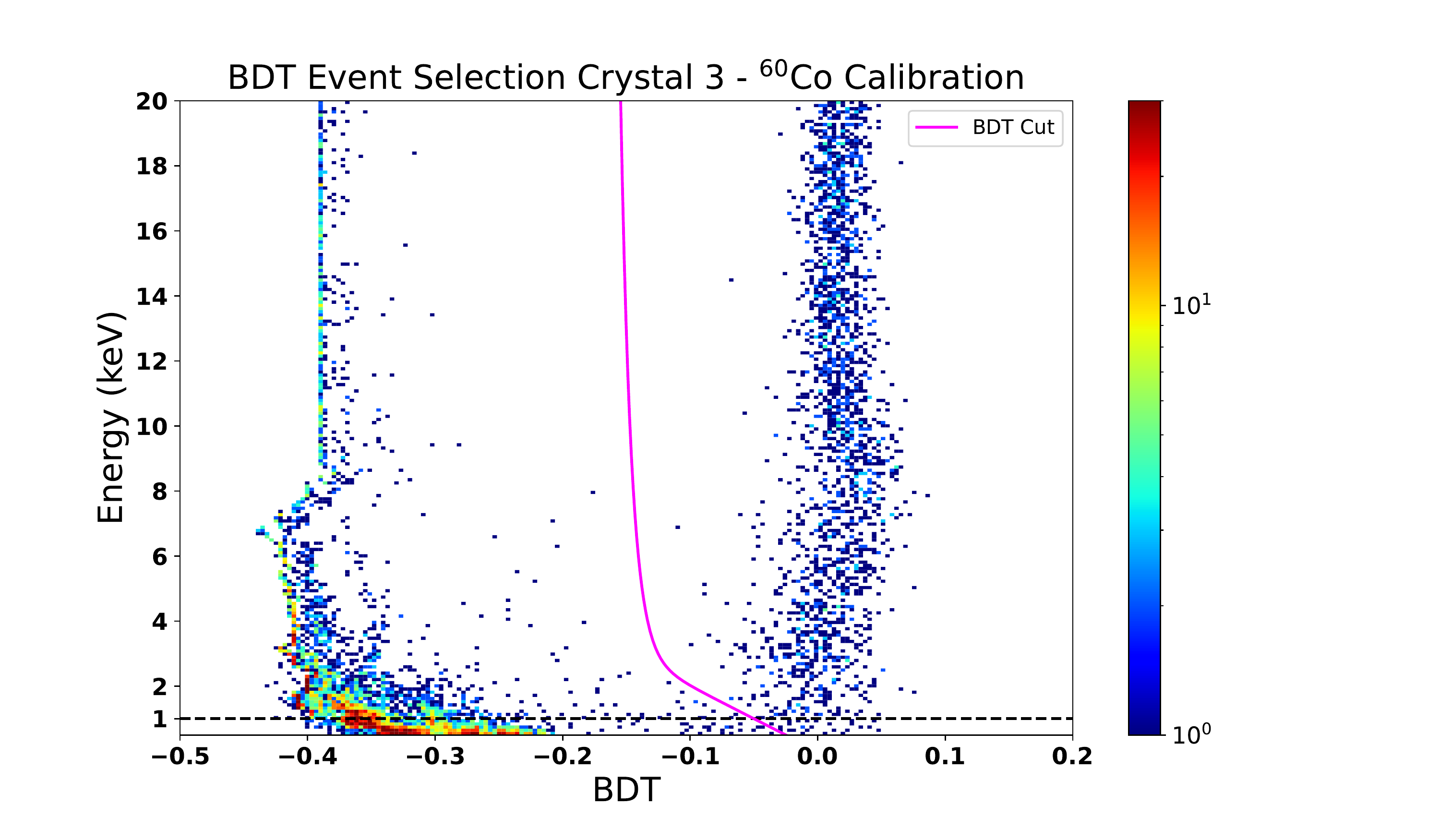}
\end{subfigure}
\begin{subfigure}(b){}
   \includegraphics[width=0.53\textwidth, trim = 20mm 25mm 20mm 10mm]{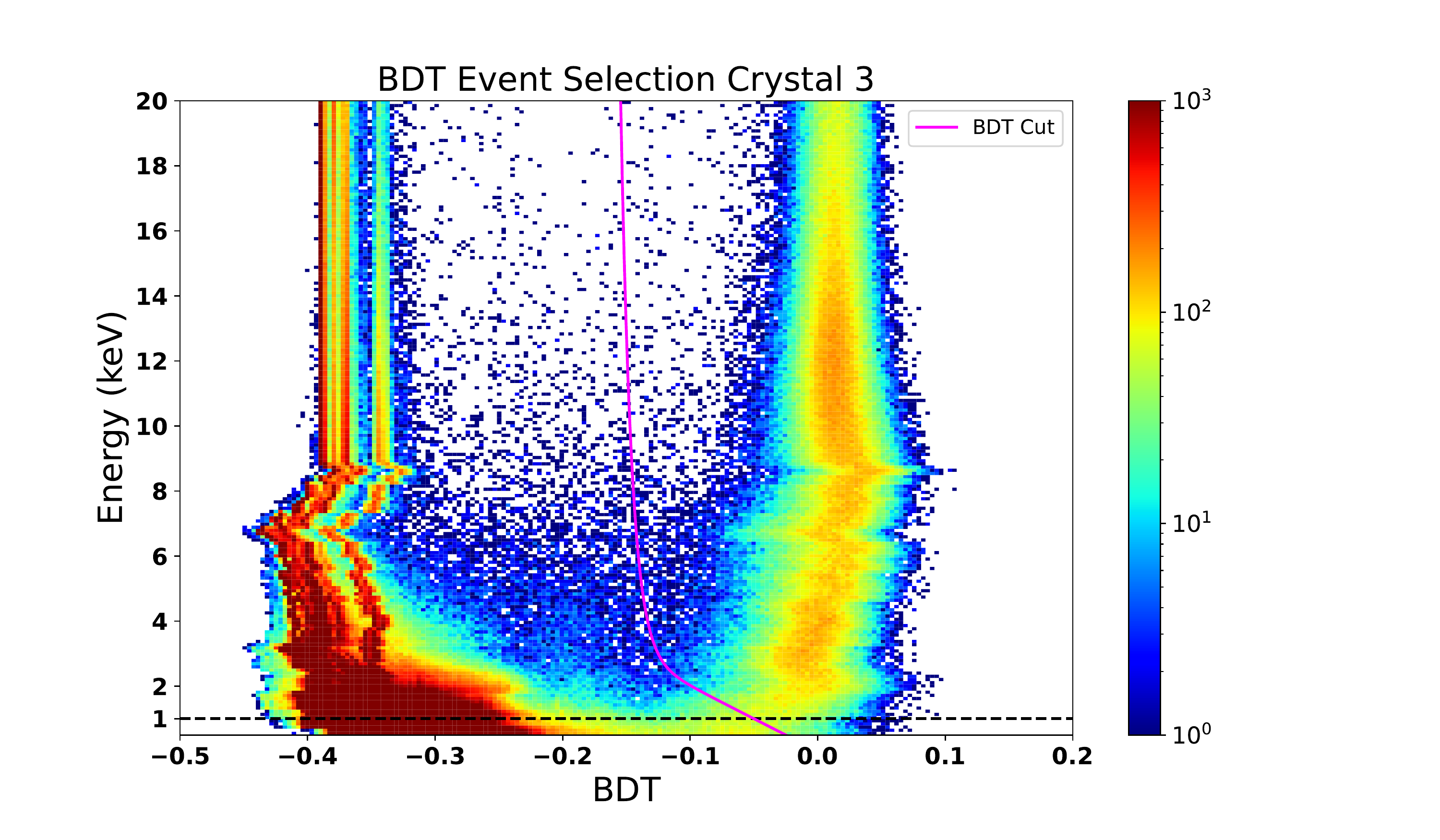}
\end{subfigure}
    \caption{``BDT'' event selection for crystal number 3. (a) Data from the detector calibration with a $^{60}$Co source. (b) Physics data taken for around 3 years. The ``BDT'' selection removes events to the left of the magenta solid line The 1 keV threshold is represented by the black dashed line.}
    \label{fig:5}
\end{center}
\end{figure}

Currently, COSINE-100 works with a threshold of 1 keV, and the noise event selection has an efficiency of around 60\% near the threshold. The main parameter used by the collaboration to reduce noise events is a Boosted Decision Tree (BDT), and is based on a machine learning algorithm, which uses the Boosted Decision Tree technique. Different crystal signal properties are used to define the ``BDT'' value for each event, such as the mean time, the energies of the first half and second half, and the amplitude of typical signals from noise and scintillations \cite{56}. Data measured in the crystals calibration with a $^{60}$Co source are used to define the cuts in the ``BDT'' parameter for each crystal. Figure \ref{fig:5} shows the ``BDT'' parameter values considering the calibration data and 2.82 years physics data set, which was analyzed in this work.

Noise events typically have low values of the ``BDT'' parameter, whilst scintillation events have higher values.

\begin{figure}[!htb]
\centering
\includegraphics[scale=0.40, trim = 0mm 7mm 15mm 0mm]{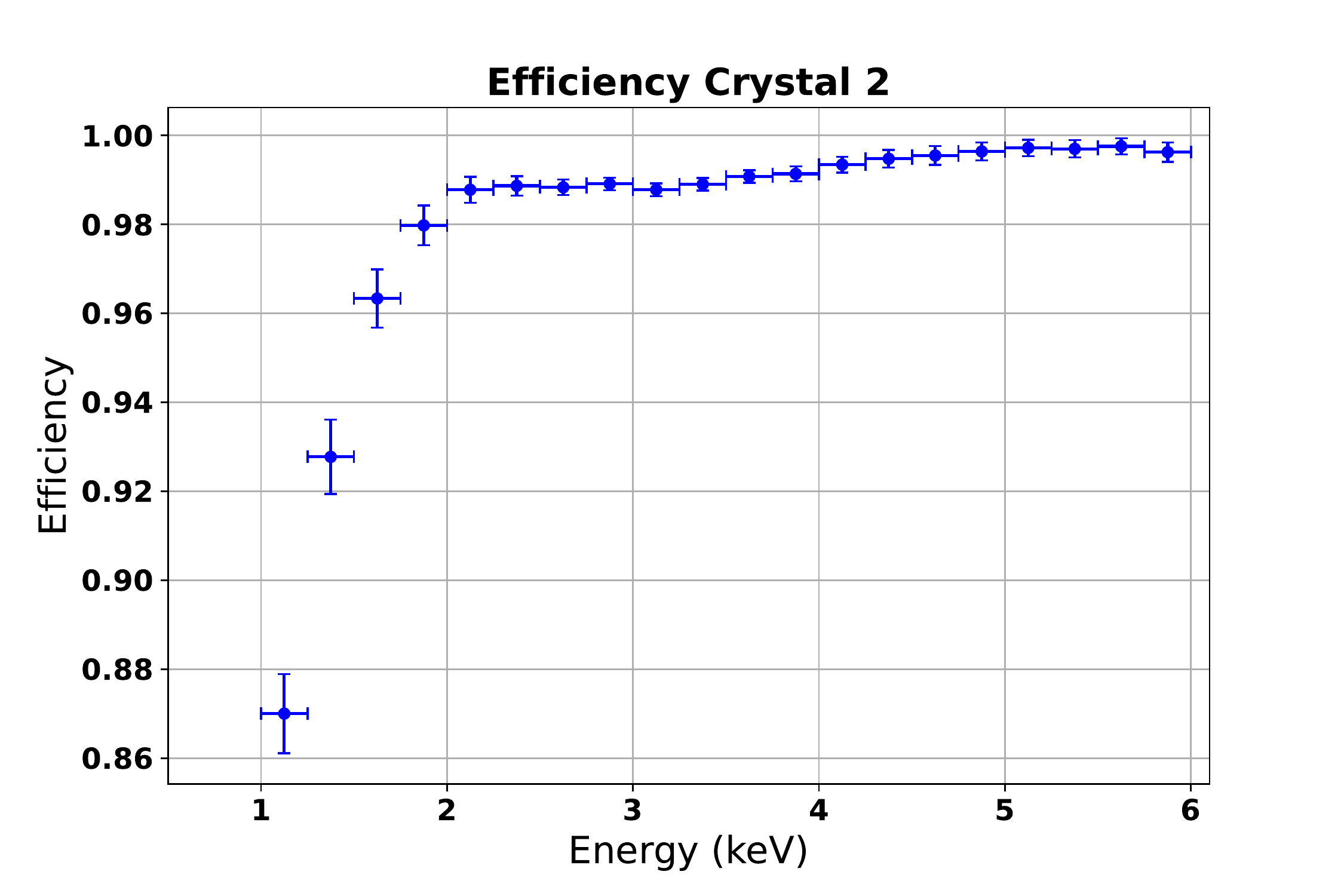}
\caption{Efficiencies for the ``BDT'' event selection in crystal 2 for energies from 1 keV to 6 keV. Above 6 keV, the detection efficiency is considered to be 100\%.}
\label{fig:6}
\end{figure}

The $^{60}$Co calibration data is used to determine the ``BDT'' event selection efficiency for each 0.25 keV energy interval in each crystal. The efficiencies for crystal number 2 are shown in Figure \ref{fig:6}.

\subsection{Annual Modulation Fit}\label{sec:analB}
The background for each crystal is well studied by the collaboration, leading to the possibility of obtaining the initial activity of each of its background components. The COSINE-100 background model consists of a component with constant background, composed of radionuclides with very long decay time, as the $^{40}$K, $^{238}$U, and $^{232}$Th nuclides, and eight components with exponential activities, represented by $^{210}$Pb, $^{121}$Te, $^{121m}$Te, $^{127m}$Te, $^{3}$H, $^{109}$Cd, $^{22}$Na, and $^{113}$Sn. Then, for each crystal, the background can be described as:
\begin{equation}\label{eq:18}
    B_n=C_n+\sum_{i=1}^{8} a_{i,n}\,e^{-\lambda_i\,t}
\end{equation}
    
\noindent where $B_n$ represents the event rate due to the background in the $n^{th}$ crystal, $C_n$ represents the event rate due to the constant background component in the $n^{th}$ crystal, and $a_{i,n}$ and $\lambda_i$ represent the initial event rate and the decay constant of the $i^{th}$ component of cosmogenics in the $n^{th}$ crystal, respectively.

Considering this background model and the expected modulation (\ref{eq:4}) for the solar dark photon and solar DFSZ and KSVZ axion models ($R \propto d^{-2}$), the event rate in each crystal can be described as:
\begin{equation}\label{eq:19}
    R_n =C_n+\sum_{i=1}^{8} a_{i,n}\,e^{-\lambda_i\,t}+A_{-2}
\end{equation}

Considering the expected modulation (\ref{eq:6}) for the KK solar axion model ($R \propto d^{-4}$), the event rate in each crystal should be:
\begin{equation}\label{eq:20}
    R_n =C_n+\sum_{i=1}^{8} a_{i,n}\,e^{-\lambda_i\,t}+A_{-4}
\end{equation}
    
\noindent where $R_n$ represents the event rate in the $n^{th}$ crystal, $A_{-2}$ and $A_{-4}$ represents the expected modulation, which is given by equations \ref{eq:4} and \ref{eq:6}, and should be the same for all crystals.

Expressions \ref{eq:19} and \ref{eq:20} were fitted to physics data of the five NaI(Tl) crystals analyzed after event selection. The background components ($C_i$ e $a_{i,n}$) were left free and independent for each crystal, and the modulation amplitudes ($A$) were left free but forced to be the same for all crystals. Furthermore, the fits were performed in energy intervals of 1 keV, in order to refine the search for the modulations, using Monte Carlo techniques based on a Bayesian analysis approach \cite{58}, similarly to the procedure adopted in the COSINE-100 WIMP annual modulation analysis \cite{59}. The posterior distribution of the activities of each background component ($C_n$, and $a_{i,n}$) and the modulation amplitude ($A$) were calculated by:
\begin{equation}
    P(A|\mathbf{D})=N\,\int d\mathbf{C} \int d\mathbf{a}~\mathcal{L}(\mathbf{D}|A,\mathbf{C},\mathbf{a})~\pi(A,\mathbf{C},\mathbf{a})
\end{equation}

\noindent where N is a normalization constant, $A$ is the modulation amplitude; $\mathbf{C}$ and $\mathbf{a}$ are vectors related to the activities of the constant and exponential components of the background, respectively; $\pi(A,\mathbf{C},\mathbf{a})$ are the prior distributions also related to the background components and the modulation amplitude; $\mathbf{D}$ represents the observed data; and $\mathcal{L}(\mathbf{D}|A,\mathbf{C},\mathbf{a})$ is the likelihood which was generated with Poissonian probabilities (P) by:
\begin{equation}
    \mathcal{L}(\mathbf{D}|A,\mathbf{C},\mathbf{a})=\prod_n^5 \prod_j^{N_{bin}^n} P(D_{nj}|E_{nj})
\end{equation}

\noindent where $N_{bin}^n$ is the number of time bins in the $n^{th}$ crystal; and $D_{nj}$ and $E_{nj}$ are the observed and expected number of events, respectively, in the $n^{th}$ crystal and $j^{th}$ time bin. $E_{nj}$ is obtained from the integration of the expected event rate ($R_n$) over the duration of the $j^{th}$ time bin. Since five crystals were analyzed, the first product has 5 terms.

The mean and statistical and systematic uncertainties considered in the priors of the activities of each background component are taken from the measured values from the collaboration background study \cite{57}.

\begin{figure}[!htb]
\centering
\includegraphics[scale=0.51, trim = 15mm 8mm 30mm 17mm]{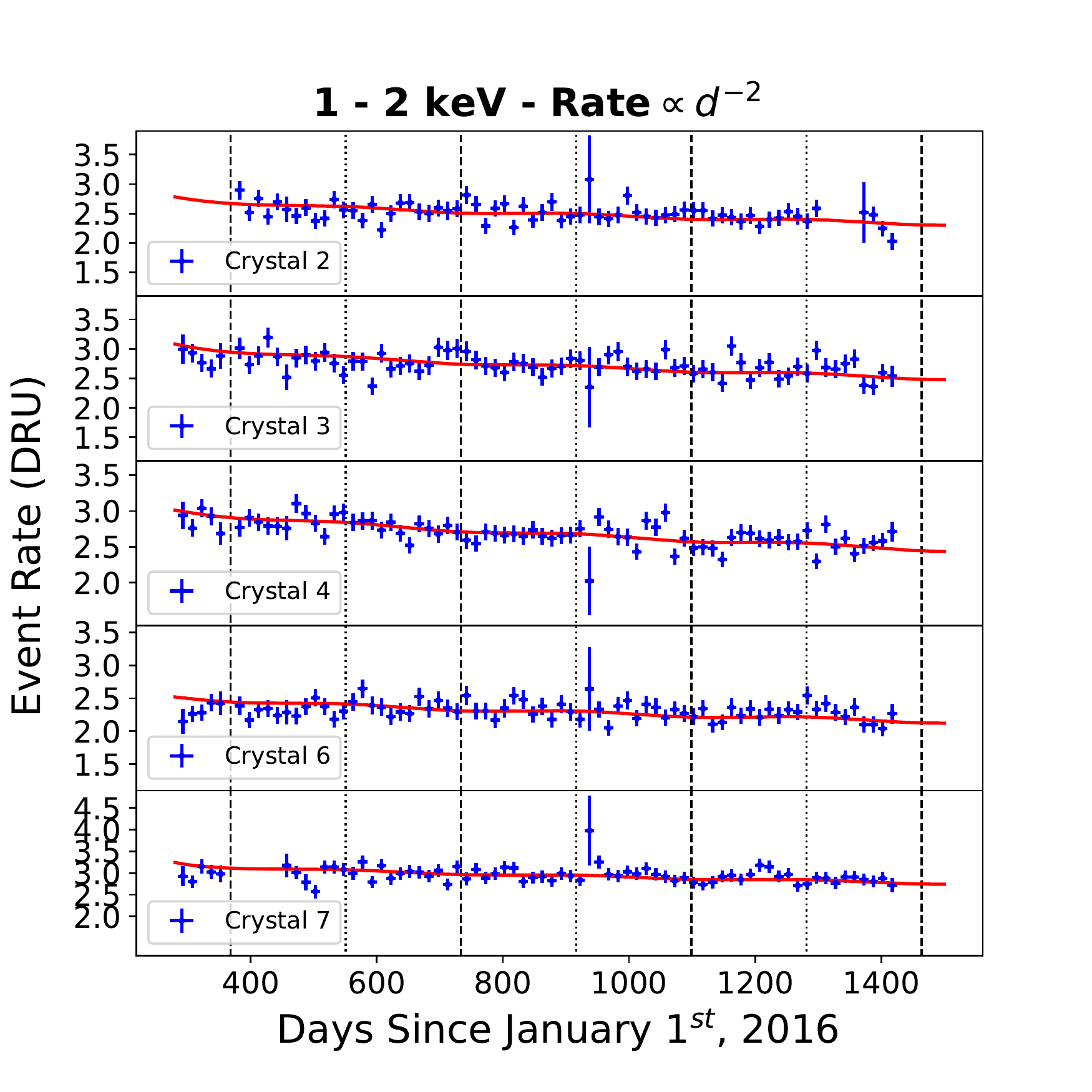}
\caption{Event rate for all crystals analyzed (blue data points), considering 15 days time bin and 1-2 keV energy range. The fit (red curves) for this energy range resulted in an amplitude of $-0.0250 \pm 0.0106$ DRU. The vertical black dotted lines refer to the aphelion dates, and the dashed lines refer to the perihelion dates.}
\label{fig:7}
\end{figure}

Figure \ref{fig:7} shows the event rate for each of the five crystals analyzed, as well as the fit of expression \ref{eq:19} to the 1-2 keV energy range. 

\begin{figure}[!htb]
\begin{center}
\begin{subfigure}(a){}
   \includegraphics[width=0.51\textwidth, trim = 0mm 0mm 10mm 0mm]{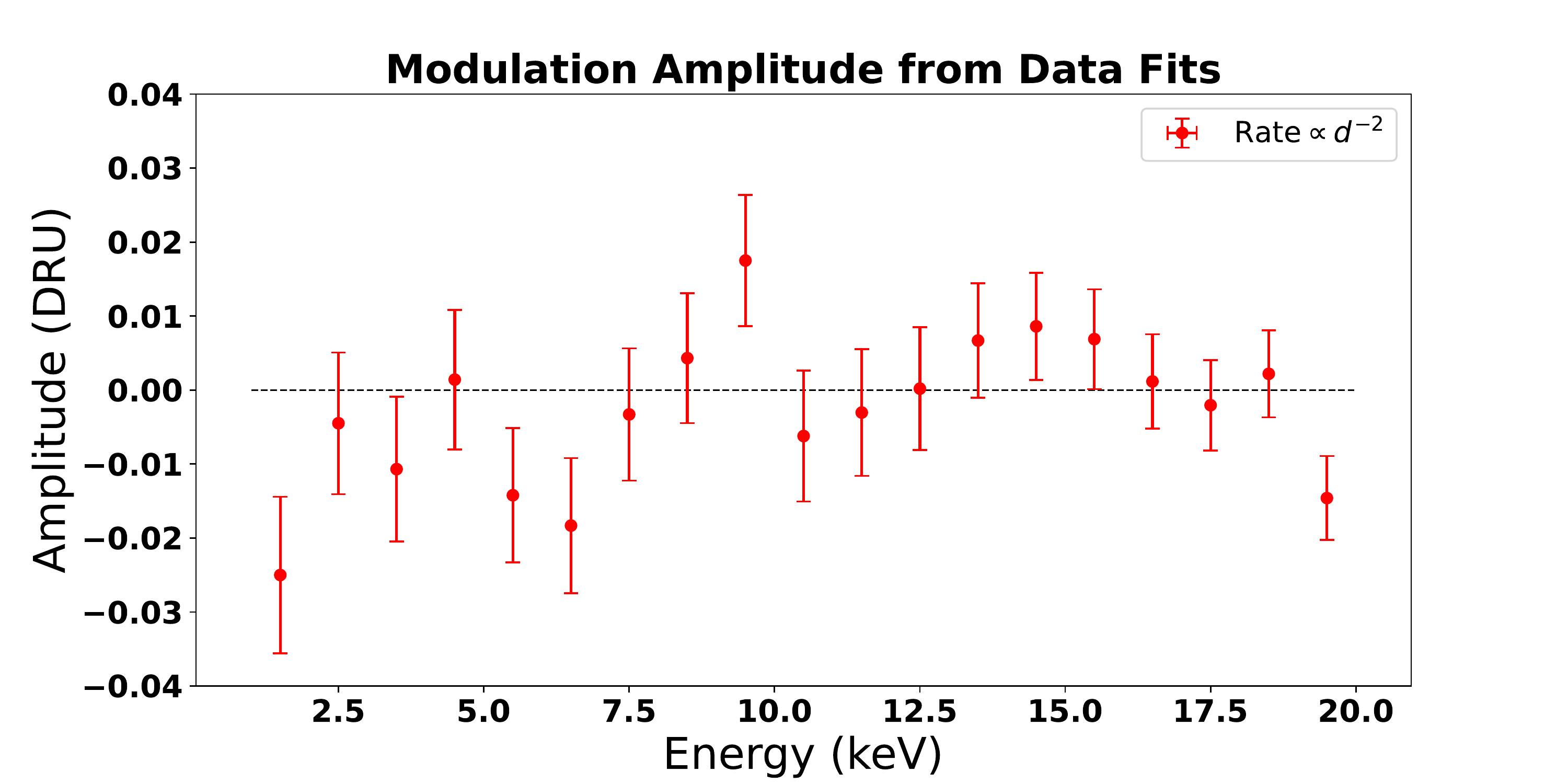}
\end{subfigure}

\begin{subfigure}(b){}
   \includegraphics[width=0.51\textwidth, trim = 0mm 20mm 10mm 0mm]{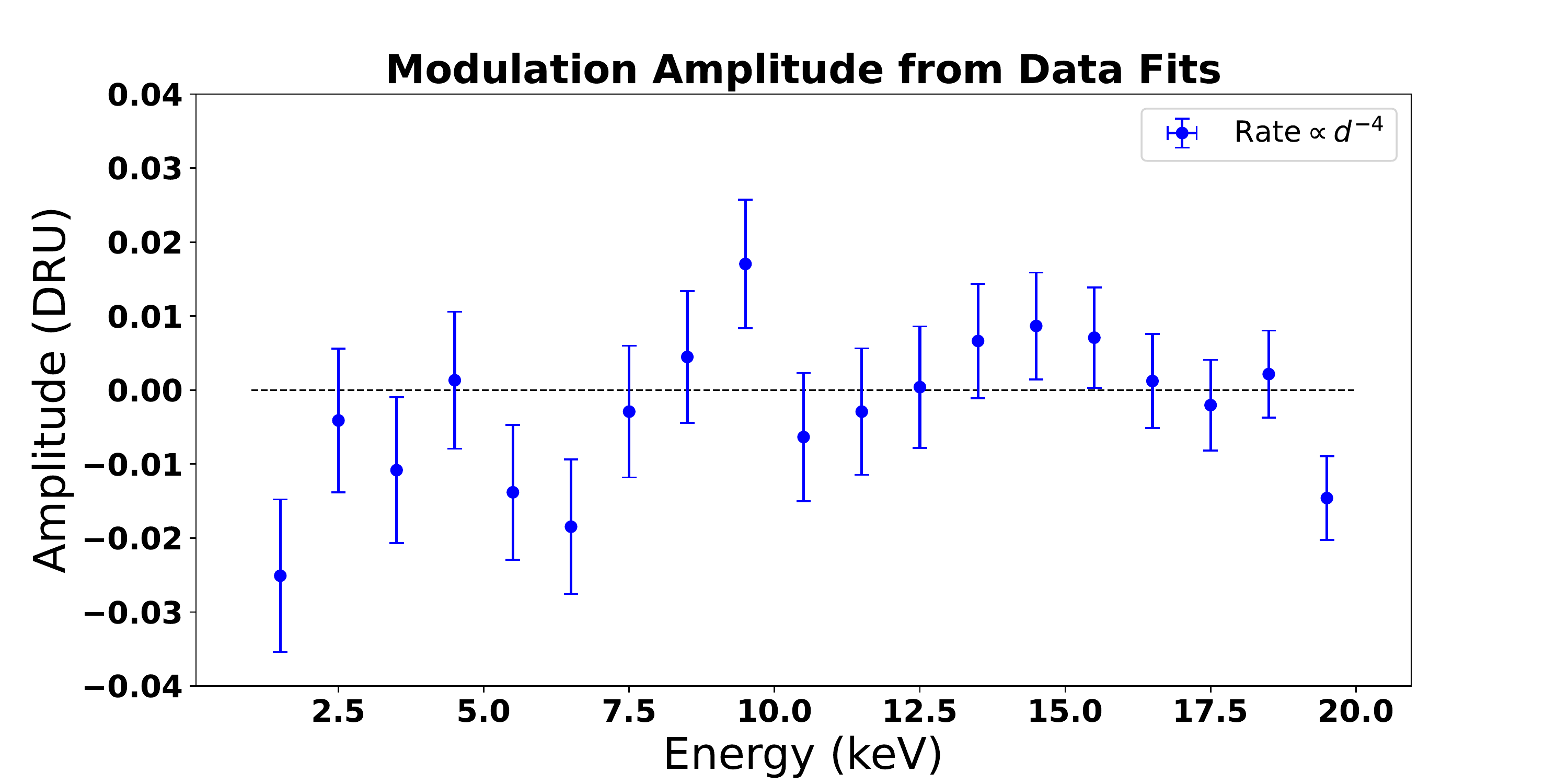}
\end{subfigure}
\caption{Amplitudes for the expected modulations simultaneous fits for each energy interval. (a) Considering the $R \propto d^{-2}$ model. (b) Considering the $R \propto d^{-4}$ model. The dashed black line represents the null modulation hypothesis.}
\label{fig:8}
\end{center}
\end{figure}

Since expressions \ref{eq:19} and \ref{eq:20} for the expected modulation are very similar, the amplitude results from the fits of both models are almost the same for all energy intervals. Figure \ref{fig:8} shows the obtained results for both expected modulation amplitudes.  

\subsection{Upper Limits}
Assuming $R \propto d^{-2}$, the expected amplitudes can be obtained from the calculated spectra for $d=a=1.496\times10^{11}$ m, according to equations \ref{eq:21} and \ref{eq:22}.

\begin{equation}
    R_{\rm{aphelion}}=\frac{R_a}{(1+e)^2}
    \label{eq:21}
\end{equation}
    
\begin{equation}
    R_{\rm{perihelion}}=\frac{R_a}{(1-e)^2}
    \label{eq:22}
\end{equation}
    
\noindent where $R_{\rm{aphelion}}$ and $R_{\rm{perihelion}}$ are the spectra when Earth is at aphelion and perihelion, respectively; and $R_a$ is the spectrum for $d=a=1.496 \times 10^{11}$ m.

The expected amplitude spectrum is then given by equation \ref{eq:23}.

\begin{equation}
    (\rm{Amplitude})=\frac{R_{\rm{perihelion}}-R_{\rm{aphelion}}}{2}
    \label{eq:23}
\end{equation}
 
\begin{figure}[!htb]
    \centering
    \includegraphics[scale=0.43, trim = 5mm 10mm 20mm 10mm]{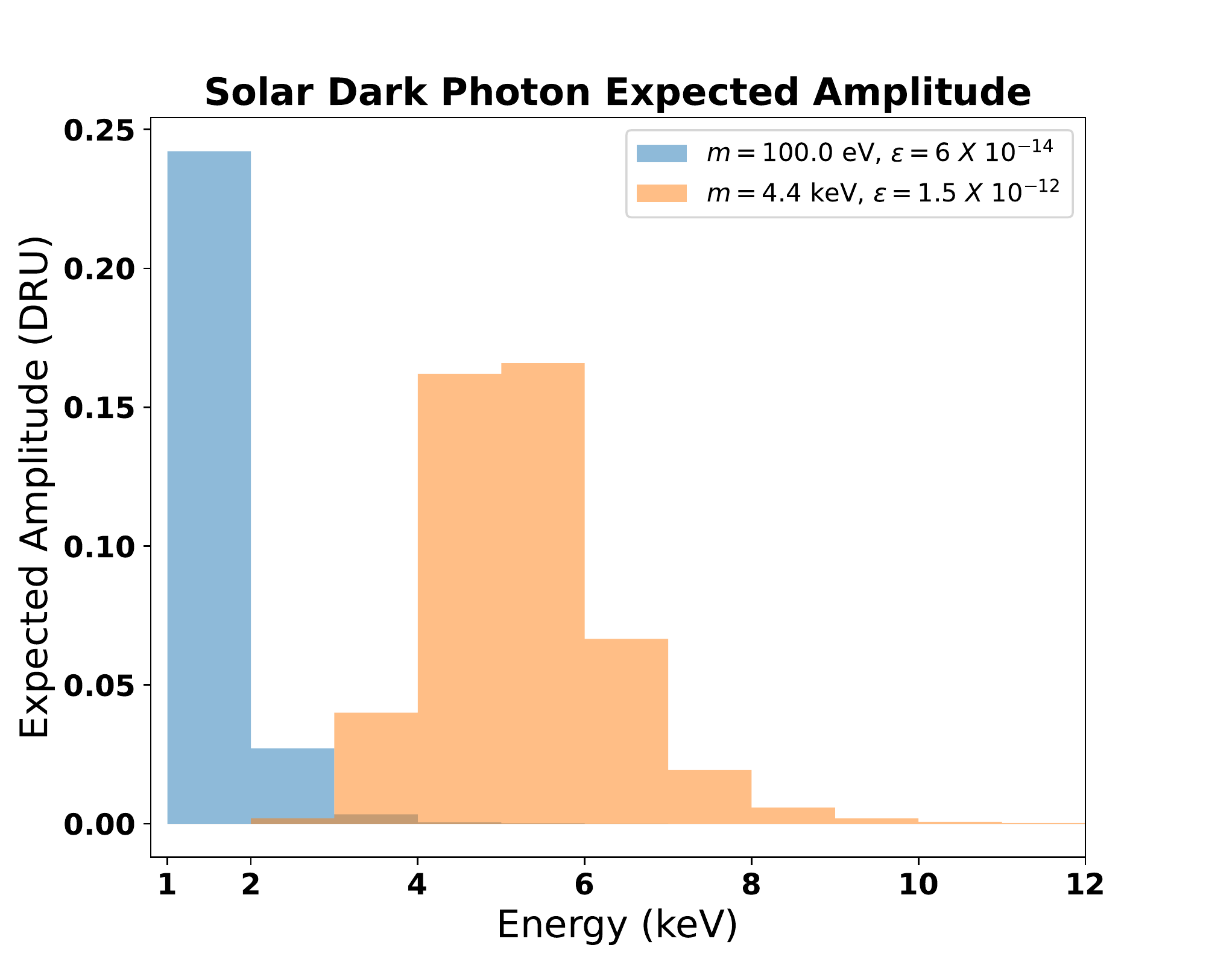}
    \caption{Amplitudes for the expected annual modulation in the event rate for solar dark photons with mass of 100 eV and $\epsilon=6 \times 10^{-14}$ (blue), and solar dark photons with mass of 4.4 keV and $\epsilon=1.5 \times 10^{-12}$ (orange). The amplitudes above 12 keV are very small, and were not shown in the plot.}
    \label{fig:9}
\end{figure}

The amplitude spectra for two different values of $\epsilon$ and mass of solar dark photons, and considering 1 keV energy bins are shown in Figure \ref{fig:9}.

\begin{figure}[!htb]
    \centering
    \includegraphics[scale=0.37, trim = 25mm 10mm 35mm 10mm]{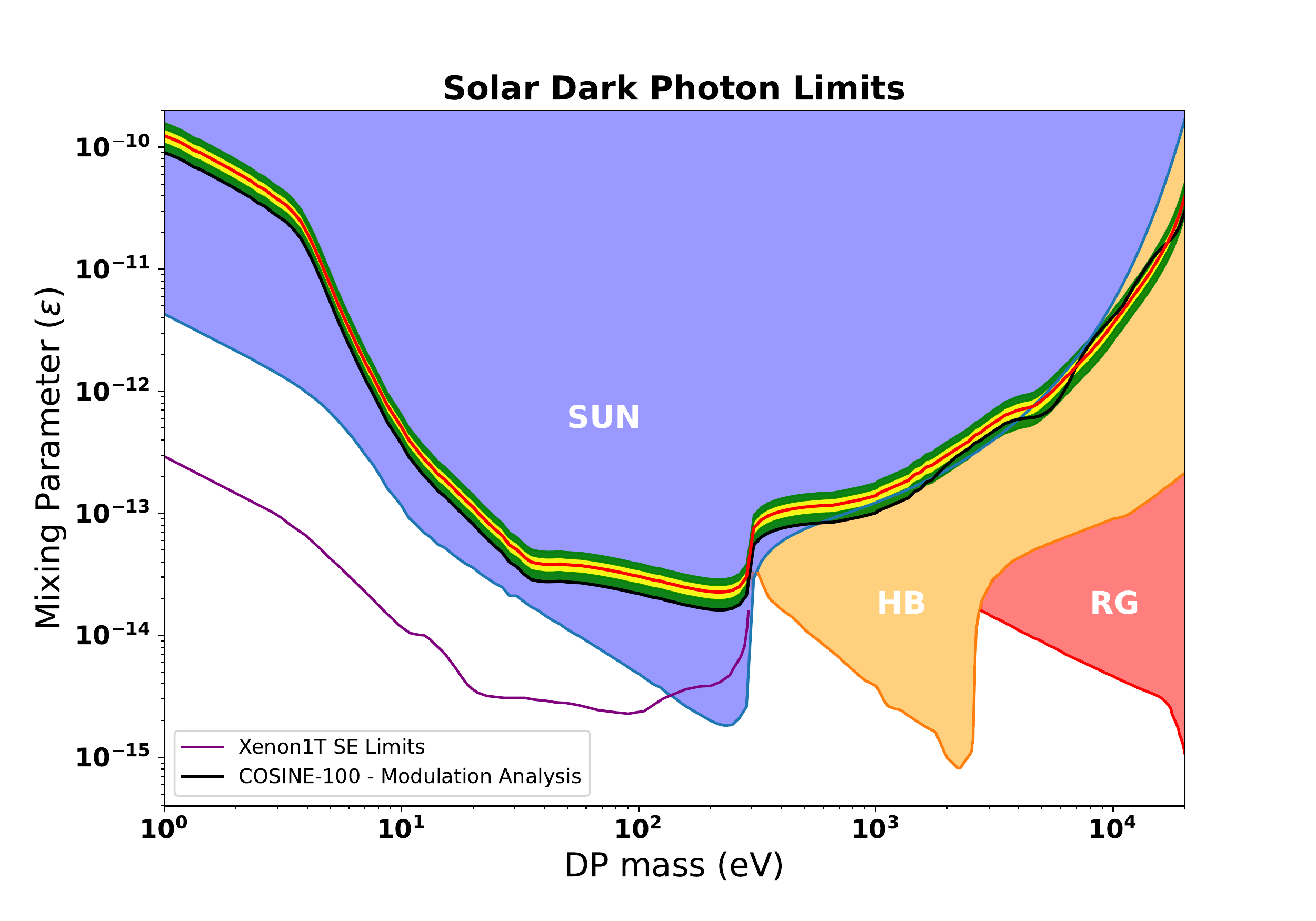}
    \caption{Exclusion plot for solar dark photons, showing the 90\% C.L. upper limits determined in this analysis. The black line is the upper limit derived from the COSINE-100 2.82 data-set analyzed. The red line shows the median of projected sensitivity. The 1 $\sigma$ and 2 $\sigma$ bands are shown by yellow and green shaded regions, respectively. Upper limits determined by Sun observations \cite{49}, red giants and horizontal branch stars studies \cite{26}, and from the XENON1T experiment \cite{61} are also shown for comparison.}
    \label{fig:10}
\end{figure}

As seen in Figure \ref{fig:8}, the observed amplitudes are all compatible with the no modulation hypothesis within 3 $\sigma$, and many of them are negative, which means a phase different from the fixed $t_0 = 3$ days. Hence, it is possible to determine upper limits for the parameters of each studied model (coupling constant or mixing parameter and particle mass) according to the Feldman-Cousins method \cite{60}. Figure \ref{fig:10} shows the 90\% C.L. upper limits for the solar dark photons model determined in this work.

\begin{figure}[!htb]
    \centering
    \includegraphics[scale=0.44, trim = 5mm 10mm 15mm 10mm]{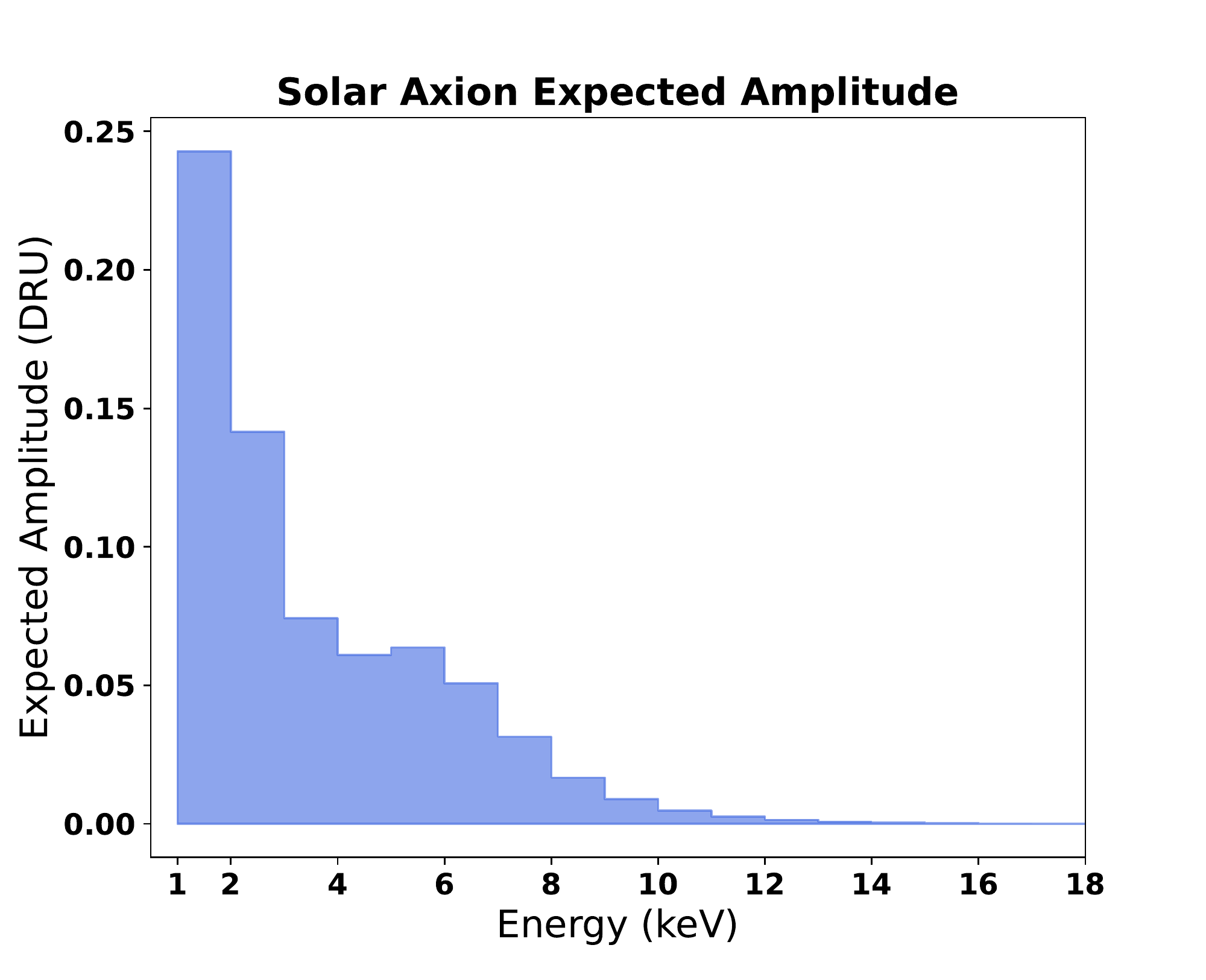}
    \caption{Expected amplitude for DFSZ and KSVZ solar axions in the COSINE-100 crystals for each 1 keV energy interval.}
    \label{fig:11}
\end{figure}

In the same way as for the solar dark photons, the event rate generated by DFSZ and KSVZ solar axions should be proportional to $r^{-2}$. Figure \ref{fig:11} shows the calculated amplitudes for the expected modulation.

\begin{figure}[!htb]
    \centering
    \includegraphics[scale=0.375, trim = 25mm 10mm 35mm 10mm]{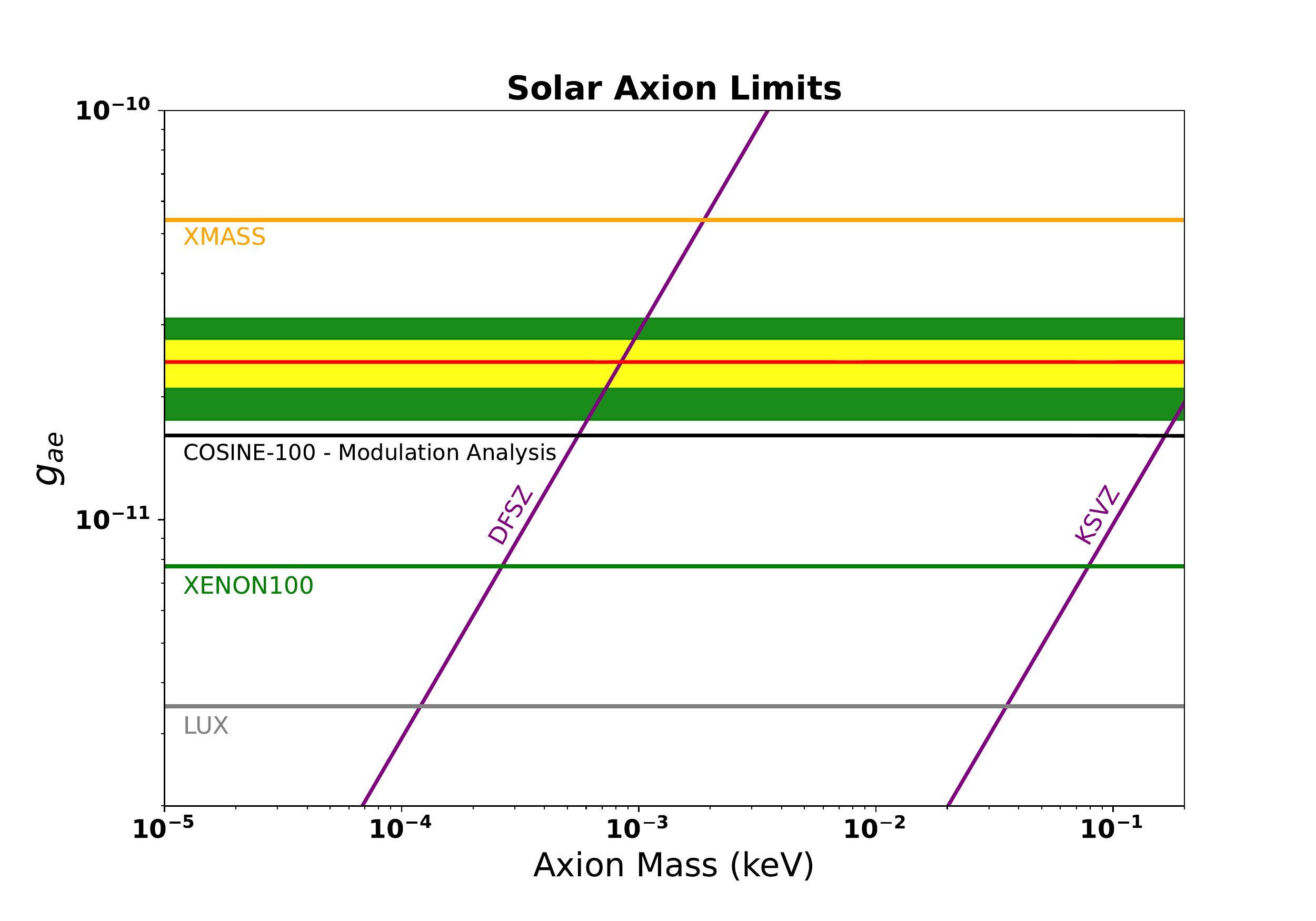}
    \caption{Exclusion plot for DFSZ and KSFZ solar axions showing the 90\% C.L. upper limits determined in this analysis. Limits from XMASS \cite{62}, XENON100 \cite{63}, LUX \cite{29}, and values theorized by DFSZ and KSVZ models are also shown for comparison. The projected sensitivity (red line) was $g_{ae}=2.49 \times 10^{-11}$, and upper limits from physics data analysis (black line) was $g_{ae}=1.69 \times 10^{-11}$.}
    \label{fig:12}
\end{figure}

As the expected modulation is the same as in the solar dark photon model, no modulation compatible with the expected modulation was observed. Figure \ref{fig:12} shows the upper limits with 90\% C.L. determined from this analysis.

For solar dark photons with mass below 2 keV, the 90\% C.L. upper limits determined in this analysis (Figure \ref{fig:10}) are in the lower $2\sigma$ band of the projected sensitivity. The same behaviour is seen for DFSZ and KSVZ solar axions (Figure \ref{fig:12}), but the limits of this analysis are below the $2\sigma$ projected sensitivity band. This behaviour is due to the modulation amplitude results from the data fit (Figure \ref{fig:8}) combined with the spectra shape of both dark matter models, especially for the energies below 8~keV. In this region, most obtained amplitudes are negative, and the spectra for both models are higher and more relevant to the upper limits determination. The combination of these two factors leads to limits better than the median of the projected sensitivity. The modulation amplitudes derived from this analysis are anti-correlated to the COSINE-100 WIMP analysis modulation amplitudes \cite{59}, as a consequence of the phase shift of both analysis.

\begin{figure}[!htb]
    \centering
    \includegraphics[scale=0.37, trim = 25mm 10mm 35mm 10mm]{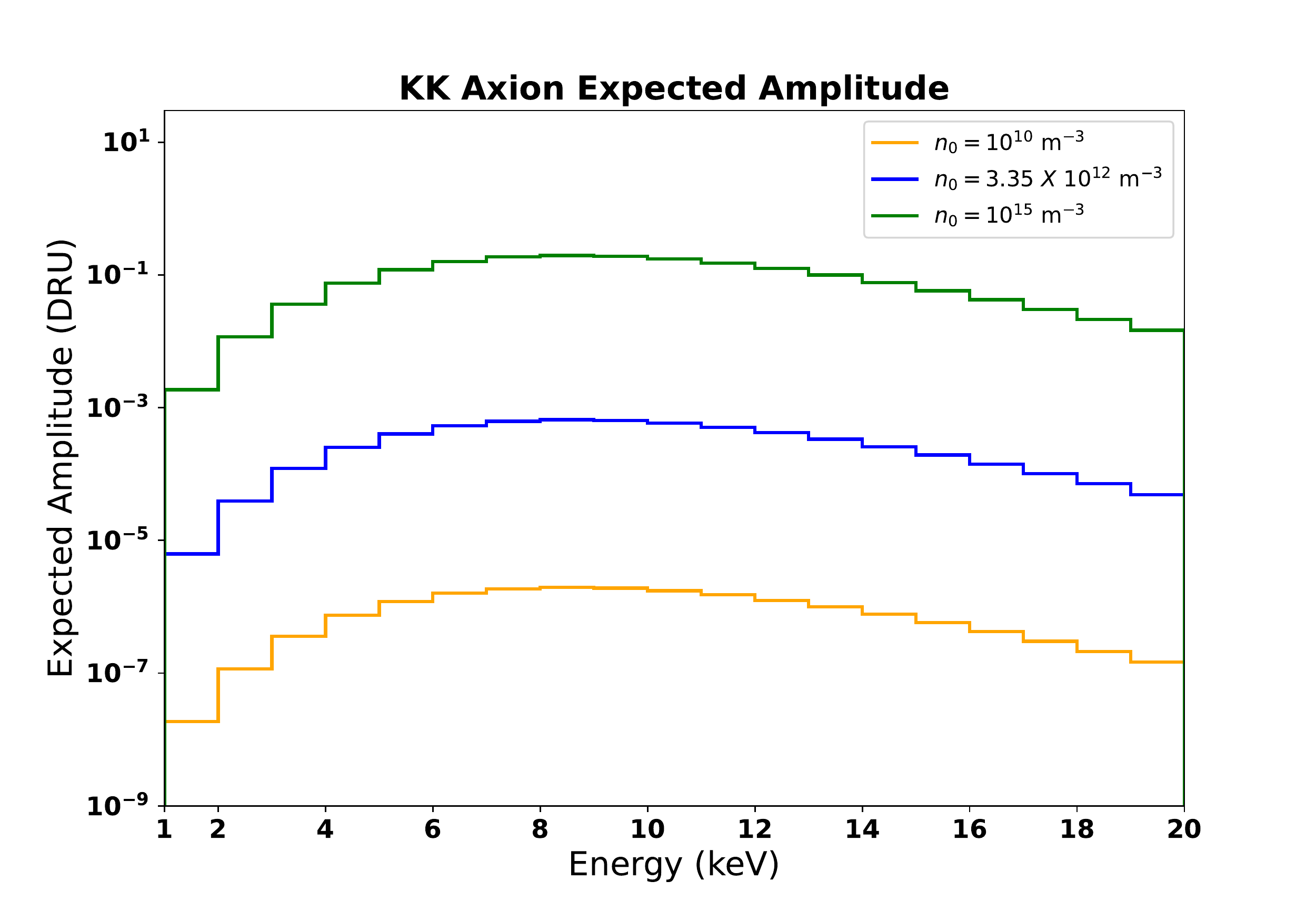}
    \caption{Expected amplitudes in the NaI(Tl) crystals, considering $g_{a \gamma \gamma}=2 \times 10^{-11}$ GeV$^{-1}$, and $n_0=10^{10}$ m$^{-3}$ (red); $n_0=3.35 \times 10^{12}$ m$^{-3}$ (blue); and $n_0=10^{15}$ m$^{-3}$ (green).}
    \label{fig:13}
\end{figure}

For KK solar axions, considering $R \propto d^{-4}$, the expected amplitudes for different numerical axion densities at Earth were calculated, from $n_0=10^{10}$ m$^{-3}$ up to $n_0=10^{15}$ m$^{-3}$, as shown in Figure \ref{fig:13}.

\begin{figure}[!htb]
    \centering
    \includegraphics[scale=0.37, trim = 25mm 10mm 35mm 10mm]{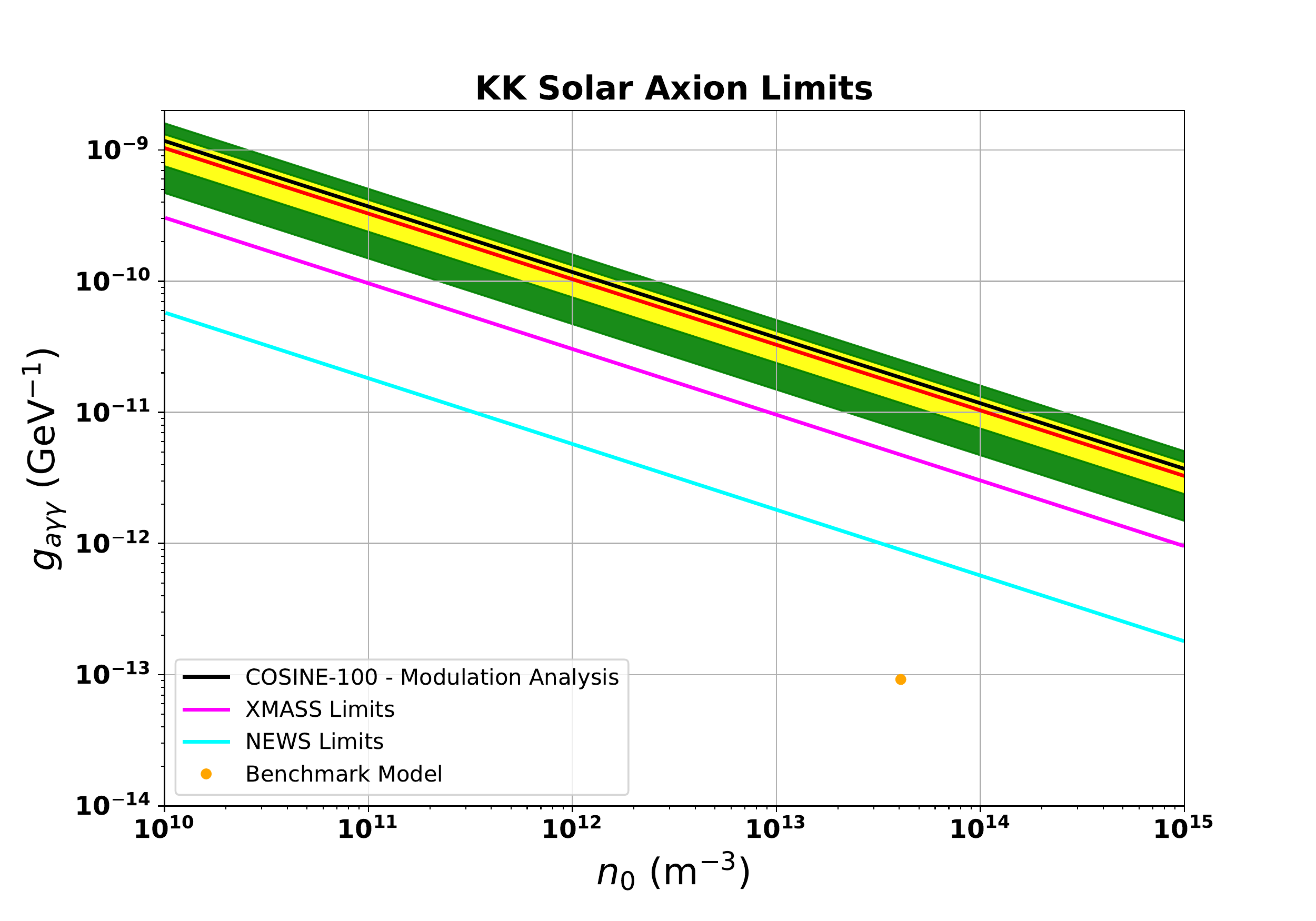}
    \caption{Upper limits with 90\% C.L. determined for the solar KK axion model from this analysis. Projected sensitivity with its 1 $\sigma$ and 2 $\sigma$ bands are shown by the red line, yellow and green bands, respectively. Upper limits from XMASS \cite{64} (magenta), NEWS \cite{65} (cyan) and the $g_{a \gamma \gamma}$ and $n_0$ values that could explain the solar coronal heating problem (orange) are also shown for comparison.}
    \label{fig:14}
\end{figure}

Figure \ref{fig:13} also shows that the observed modulation amplitudes were not compatible with the expected modulation amplitudes. Figure \ref{fig:14} shows the upper limits determined from this analysis.

\section{Sensitivity for COSINE-200}
\label{sec:sens}
The COSINE collaboration is planning to upgrade the COSINE-100 detector to an experiment with 200 kg of NaI(TI) crystals, known as COSINE-200. The research on producing crystals with reduced background and higher light yield has resulted in NaI(Tl) crystals with less contamination from internal $^{40}$K and $^{210}$Pb when compared to COSINE-100 crystals. Also, a light yield of 22 NPE/keVee (NPE is the number of photoelectrons) has already been achieved, higher than approximately 15 NPE/keVee of COSINE-100 crystals. 

\begin{figure}[!htb]
    \centering
    \includegraphics[scale=0.38, trim = 25mm 10mm 35mm 10mm]{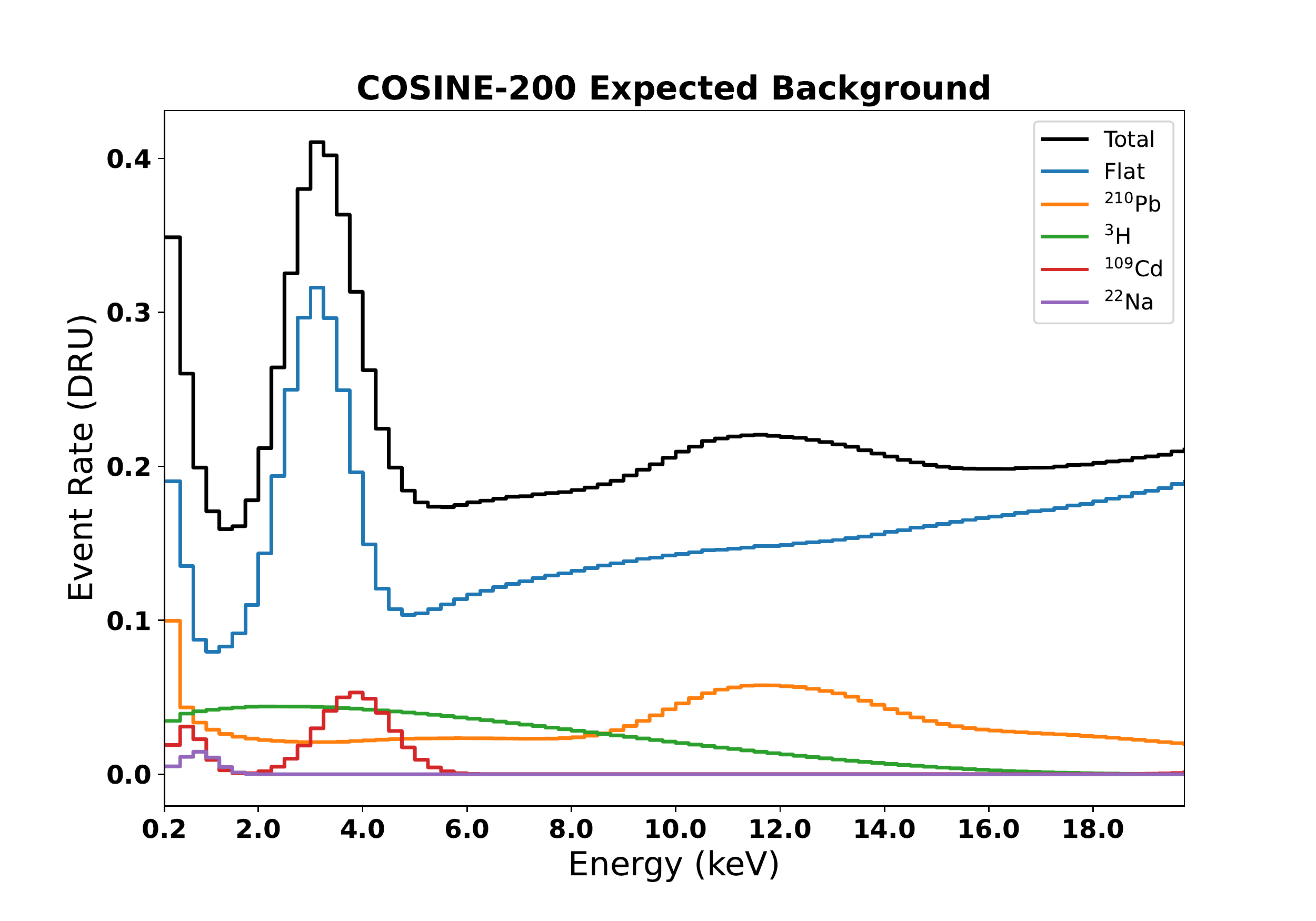}
    \caption{Background model for COSINE-200 crystals used for our sensitivity study, based on our expectation for COSINE-200. Contribution from each component is show in colored lines. Contribution from Te series and $^{113}$Sn are expected to be very small, and can be neglected.}
    \label{fig:15}
\end{figure}

Due to improvements in the crystals and in the methods capable of discriminating scintillation events from noise events with good efficiency, an analysis threshold of 5 NPE is expected for COSINE-200, meaning an energy threshold of 0.2 keV. Additionally, the background activity is expected to be lower than 0.5 DRU, as shown in Figure \ref{fig:15}. As discussed in Ref. \cite{66}, this expected background is based on tests with small crystals.

\begin{figure*}[!htb]
    \begin{center}
    \includegraphics[width=0.97\textwidth, trim = 10mm 10mm 10mm 0mm]{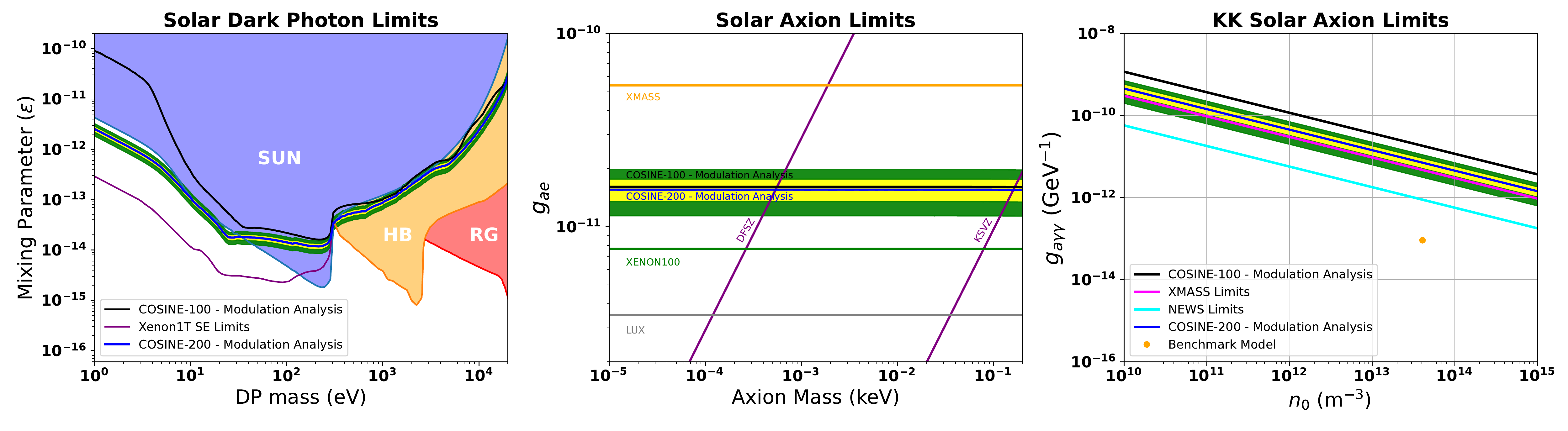} 
    \caption{Projected sensitivity for the solar dark photon (left); DFSZ and KSVZ solar axion (center); and KK solar axion (right) models, considering future COSINE-200 experiment NaI(Tl) crystals.}
    \label{fig:16}
    \end{center}
\end{figure*}

Based on the expected background spectrum, it is possible to project COSINE-200 sensitivity for the analysis performed in this work. Similarly to the procedure adopted in the analysis for COSINE-100, 1000 pseudo-data sets were generated based on the expected COSINE-200 background. Two years data exposure with 200 kg of NaI(Tl) crystals were considered. Figure \ref{fig:16} shows the projected COSINE-200 sensitivity for the three models studied in this work. 

Considering the solar dark photon model, the projected sensitivity show a considerable improvement compared to COSINE-100 physics data upper limits for dark photon masses below approximately 40 eV. However, for higher masses, the improvement is not as good. This behavior is due to the threshold reduction from 1 keV to 0.2 keV. For lower masses, the solar dark photon spectrum for energies below 1 keV is higher than for energies above 1 keV. For masses below approximately 30 eV, it should be possible to begin probing regions unexplored by the solar luminosity limits. 

Considering the DFSZ and KSVZ solar axion and the KK solar axion models, the projected sensitivity improvement comes mostly from the background reduction and increasing total crystal mass, since the spectra for these models is very low for energies below 1 keV.

\section{Conclusion}
\label{sec:concl}
A search for the expected solar dark matter annual modulation has been performed with a dataset of 2.82 years using 61.3 kg of NaI(Tl) crystals of the COSINE-100 experiment. Even though this search method is general, and could be applicable to any solar dark matter model, for specificity, three bosonic dark matter models were considered in the analysis: solar dark photons; DFSZ and KSVZ solar axions; and Kaluza-Klein solar axions. In this annual modulation analysis, amplitudes not compatible with the null modulation hypothesis and compatible with the expected modulations for each of the three studied models were not found. Consequently, upper limits for the kinetic mixing parameter and mass of the solar dark photon, for the coupling constant between axions and electrons and mass of the solar DFSZ and KSVZ axion, and for the coupling constant between axions and two photons and mass of the solar KK axions were determined. The most constraining limits for the models exclude solar dark photons with mixing parameter above $1.61 \times 10^{14}$ for a mass of 215 eV; DFSZ and KSVZ solar axions with axion-electron coupling above $1.61 \times 10^{-11}$ for an axion mass below 0.2 keV; and Kaluza-Klein axions with axion-photon coupling above $1.83 \times 10^{-11}$ GeV$^{-1}$ for an axion number density of $4.07 \times 10^{13}$ cm$^{-3}$. For each of the three models, the obtained upper limits are less stringent than limits from other experiments or astrophysical observations that were based on investigations of distinct phenomena, with different model-dependence. Projected sensitivity for two years data set of the future COSINE-200 experiment will provide an appreciable improvement for low solar dark photon masses due to reduction of the detector threshold from 1 keV to 0.2 keV. Examination of regions unexplored by solar luminosity limits should begin to be possible with the COSINE-200 detector. Nevertheless, this is the first search for an annual modulation for solar dark photons and solar DFSZ and KSVZ axions, exploiting a novel search method for these two solar dark matter particles.

\acknowledgments
We thank the Korea Hydro and Nuclear Power (KHNP) Company for providing underground laboratory space at Yangyang and the IBS Research Solution Center (RSC) for providing high performance computing resources. 
This work is supported by:  the Institute for Basic Science (IBS) under project code IBS-R016-A1, NRF-2021R1A2C3010989 and NRF-2021R1A2C1013761, Republic of Korea;
NSF Grants No. PHY-1913742, DGE-1122492, WIPAC, the Wisconsin Alumni Research Foundation, United States; 
STFC Grant ST/N000277/1 and ST/K001337/1, United Kingdom;
Grant No. 2021/06743-1 and 2022/12002-7 FAPESP, CAPES Finance Code 001, CNPq 131152/2020-3, Brazil.

\bibliographystyle{ieeetr}
\bibliography{ref.bib}
\end{document}